\documentclass[aps,nofootinbib, prd, amsmath, floats, floatfix, twocolumn,superscriptaddress, showpacs,reprint]{revtex4-1}

\usepackage{graphicx}
\usepackage{amsmath,amssymb}
\usepackage{amsfonts}
\usepackage{xspace} 
\usepackage[usenames]{color}
\usepackage{dcolumn}
\usepackage{bm}
\usepackage{mathrsfs}
\usepackage[colorlinks=true]{hyperref}
\usepackage[all]{hypcap} 
\usepackage[utf8]{inputenc} 
\usepackage{bbold}
\usepackage{algorithm}
\usepackage[noend]{algpseudocode}

\usepackage{etoolbox}
\usepackage{tikz}
\usetikzlibrary{tikzmark}
\usetikzlibrary{calc}

\usepackage{tabularx}

\def\be{\begin{equation}}
\def\ee{\end{equation}}
\def\beq{\begin{eqnarray}}
\def\eeq{\end{eqnarray}}

\newcommand{\tn}{\textnormal}

\begin{document}

\title{Solving the relativistic inverse stellar problem through gravitational waves observation of binary neutron stars}

\author{Tiziano Abdelsalhin}\email{tiziano.abdelsalhin@roma1.infn.it}
\affiliation{Dipartimento di Fisica, Sapienza Università di Roma \& Sezione INFN Roma1, P.A. Moro 5, 00185, Roma, Italy}
\author{Andrea Maselli}\email{andrea.maselli@tecnico.ulisboa.pt}
\affiliation{CENTRA, Departamento de Fisica , Instituto Superior Tecnico - IST, 
Universidade de Lisboa, - UL, Avenida Rovisco Pais 1, 1049 Lisboa, Portugal}
\author {Valeria Ferrari}\email{valeria.ferrari@roma1.infn.it}
\affiliation{Dipartimento di Fisica, Sapienza Università di Roma \& Sezione INFN Roma1, P.A. Moro 5, 00185, Roma, Italy}


\date{\today}
\begin{abstract}
The LIGO/Virgo collaboration has recently announced the direct
detection of gravitational waves emitted in the coalescence of a
neutron star binary.
This discovery allows, for the first time,  to set  new
constraints on the behavior of matter at supranuclear density,
complementary with those coming from astrophysical observations in the
electromagnetic band. In this paper we demonstrate the feasibility of
using gravitational signals to solve the relativistic inverse stellar
problem, i.e., to reconstruct the parameters of the equation of state
(EoS) from measurements of the stellar mass and tidal Love number. We
perform Bayesian inference of mock data, based on different models of
the star internal composition, modeled through piecewise polytropes.
Our analysis shows that the detection of a small number of sources
by a network of advanced interferometers would allow to put accurate
bounds on the EoS parameters, and to perform a
model selection among the realistic equations of state  proposed
in the literature.  \end{abstract}

\maketitle 

\section{Introduction}
Decades of experimental and theoretical efforts have finally led
gravitational wave astronomy to emerge as a new field of research and
an extraordinary lookout on the high energetic phenomena of our
Universe. After the first binary black hole coalescence observed by
LIGO \cite{LIGO1,PhysRevLett.116.241103,PhysRevLett.118.221101},
Advanced Virgo has recently joined the quest, leading to the discovery
of a further double black hole system, greatly improving the
capability to localize the source \cite{PhysRevLett.119.141101}. More
recently, the LIGO/Virgo collaboration has announced the first
detection of a gravitational wave (GW) signal associated to the
inspiral of two coalescing neutron stars
\cite{PhysRevLett.119.161101}. GW170817 also is the first
astrophysical event observed simultaneously in both the gravitational
and electromagnetic bands: a short gamma ray burst detected in
coincidence has marked the dawn of the multimessenger astronomy
\cite{2041-8205-848-2-L12}.

At design sensitivity, LIGO and Virgo are expected to detect almost
one binary neutron star (NS) per week. This incoming flood of data will be extremely precious to
test gravity in a strong field regime, and to investigate the behavior
of matter in extreme conditions. Neutron star cores are
characterized by densities which may exceed the nuclear
saturation point $\rho_0 \sim 2.7 \times \text{10}^{\text{14}}\text{
g/cm}^{\text{3}}$.  A rigorous and comprehensives description of
matter in this regime is currently unavailable. The lack of
experimental data and the complexity of modeling two and three-body
strong interactions at $\rho\gg \rho_0$ leads to large uncertainties
on the star equation of state.  Various theoretical approaches have
been developed so far, that predict different scenarios for the
nuclear matter, including mixture of $npe\mu$ in $\beta$-equilibrium,
hyperon production, meson condensates and phase transitions to
deconfined quarks \cite{0004-637X-550-1-426}.

It is known that the equation of state of matter 
$p=p(\epsilon)$, which links pressure and energy density,
can be mapped to a mass-radius relation using the equation of stellar structure
\cite{1992ApJ...398..569L}; these link the microscopic properties
of matter to the macroscopic properties of the star, the radius $R$ and the mass
$M$.
A complete knowledge of the $M(R)$ profile
could, in principle, be inverted in order to determine the EoS.
However, observational uncertainties, and a limited number of
simultaneous measurements of NS observables, are the largest
obstacles in solving the {\it relativistic inverse stellar problem}
to constrain the nuclear equation of state
\cite{1992ApJ...398..569L}. 
 
In the last years, gravitational waves from binary NSs or
NS-black hole mergers have been invoked as precious source of
information, possibly able to shed new light on the star internal
composition
\cite{Hinderer:2009ca,Damour:2012yf,Maselli:2013rza,Lackey:2013axa,Lackey:2011vz,PhysRevD.91.043002,PhysRevD.89.103012}.
In this regard, the recent detection of the GW170817 event provides a
striking confirmation of the new prospects offered by GW astronomy
\cite{PhysRevLett.119.161101}.  The imprint of the equation of state
on the gravitational signal shows up during the last phases of the
inspiral\footnote{A footprint of the star EoS is also present in
the postmerger signal (see for example
\cite{0264-9381-33-8-085003,Bose:2017jvk} and references therein). The
analysis of this part of the waveform is however outside the scope of
this paper.}, when tidal interactions play a significant role and are
strong enough to induce a quadrupolar deformation in the NSs, that
affects the phase of the emitted waveform. In the adiabatic
approximation, the induced quadrupole moment $Q_{ij}$ is proportional
to the external tidal field $\mathcal{E}_{ij}$, $Q_{ij} = -\lambda
\mathcal{E}_{ij}$, where $\lambda$ is the NS \emph{tidal deformability},
which encodes the deformation properties of the star
\cite{Hinderer:2007mb,Damour:2009vw,Binnington:2009bb}.  For a given
compactness $M/R$, $\lambda$ depends on the equation of state only.
Therefore, binary mergers containing at least one NS offer a new
possibility to constrain the EoS of matter at supranuclear densities. GW170817 has
already set interesting constraints on the tidal deformability
\cite{PhysRevLett.119.161101}, which  favour low stellar
compactness and hence soft matter, in agreement with astrophysical
measurements in the electromagnetic bandwidth \cite{Ozel:2016oaf}. 

In principle, multiple observations of isolated and binary NSs
may provide a collection of pairs  $[M,R]$ or $[M,\lambda]$, dense {\it and}
accurate enough to map the correct EoS. However, error bars on these
quantities are still large, especially on $R$ and $\lambda$, and
the proposed EoSs depend on several parameters arising  on
the way hadron interactions are modeled and  on the particle
content. 
Phenomenological parametrizations of the NS equation of state
provide an effective approach to solve the inverse stellar problem
\cite{Lindblom2014}, since they allow to describe a large
class of EoSs through a relatively small set of coefficients,
to be constrained by astrophysical data. These EoSs can then
be used to combine measurements of different NS
parameters, exploiting the results of  gravitational and
electromagnetic surveys
to obtain genuine multi-wavelength constraints on the EoS
\cite{paper2}. Moreover, it may be possible that the {\it true}
equation of state differs from the models proposed in literature so
far. In this case, a phenomenological approach would be extremely
useful to constrain the main features of the correct EoS.

Phenomenological models developed so far include: (i) a spectral
representation in terms of the enthalpy, proposed by Lindblom and
collaborators \cite{lindblom2010,PhysRevD.86.084003}, (ii) the
piecewise polytropic equation of state developed by Read et al. in
\cite{read2009}, (iii) the model described by Steiner \cite{Steiner:2010fz}, 
in which the EoS is divided into four density regimes: a fixed crust below the 
nuclear saturation point $\rho_0$, one pressure-energy relation depending 
on nuclear physics parameters (as symmetry energy and proton/electron 
fraction) for $\rho\simeq \rho_0$, and two polytropic relations at larger 
densities to fit the inner core.  

In this paper we show how  GW signals emitted  by coalescing
binary neutron stars can be used to solve the relativistic inverse stellar problem, 
and to infer the EoS parameters of the piecewise polytropic parametrization.
We generate data of masses and tidal deformabilities
for two classes of EoSs which span a large range of compactness,
determining the minimum number of observations needed to fully
constrain the EoS.  We perform a Bayesian analysis by considering a
network of advanced interferometers, composed by LIGO, Virgo
and the upcoming KAGRA.  Our results suggest that few GW detections
may already be able to set accurate constraints on some of the
piecewise parameters, that can be used to make model selection among
various realistic EoS.  

The plan of the paper is the following: In Sec.~\ref{Sec:EOS} we
describe the main properties of the phenomenological parametrization
used to model the NS equation of state, and the classes of EoS considered in
this paper. In Secs.~\ref{Sec:MCMC}-\ref{Sec:setup} we outline the
numerical approach developed to constrain the parameters of the
EoS  from GW observation.  The results of our
analysis are presented in Sec.~\ref{Sec:results1}, 
where we also show how the
reconstructed parameters can be used to discriminate among different
EoSs. Concluding remarks are summarized  in Sec.~\ref{Sec:summary}.

Throughout the paper we use geometrized units, in which $G=c=1$.

\section{The equation of state}\label{Sec:EOS}
As already mentioned in the introduction, a complete and detailed
description of NS matter at supranuclear densities is still missing.
This uncertainty yields different mass-radius relations, corresponding
to distinct {\it realistic} EoS, to be constrained by observational
data. These  will ultimately allow to characterize the behavior of
nuclear forces at $\rho\gg \rho_0$ and to identify the correct
approach.  However, the complexity of such models has motivated the
quest for phenomenological frameworks, which capture the main features
of the behavior of nuclear matter and can reproduce tabulated EoS.
Among the various models proposed so far, we focus on  piecewise
polytropic equations of state in its original formulation
\cite{read2009}, although some variations\footnote{In this regard, 
Raithel \emph{et al.} have recently shown that NS masses and radii measured by electromagnetic surveys 
may be exploited to reconstruct some features of the parametrized EoS 
\cite{2017ApJ...844..156R,2016ApJ...831...44R}.} have been also considered in literature \cite{Ozel:2009da}. 

\begin{figure}[th]
\centering
\includegraphics[width=6cm]{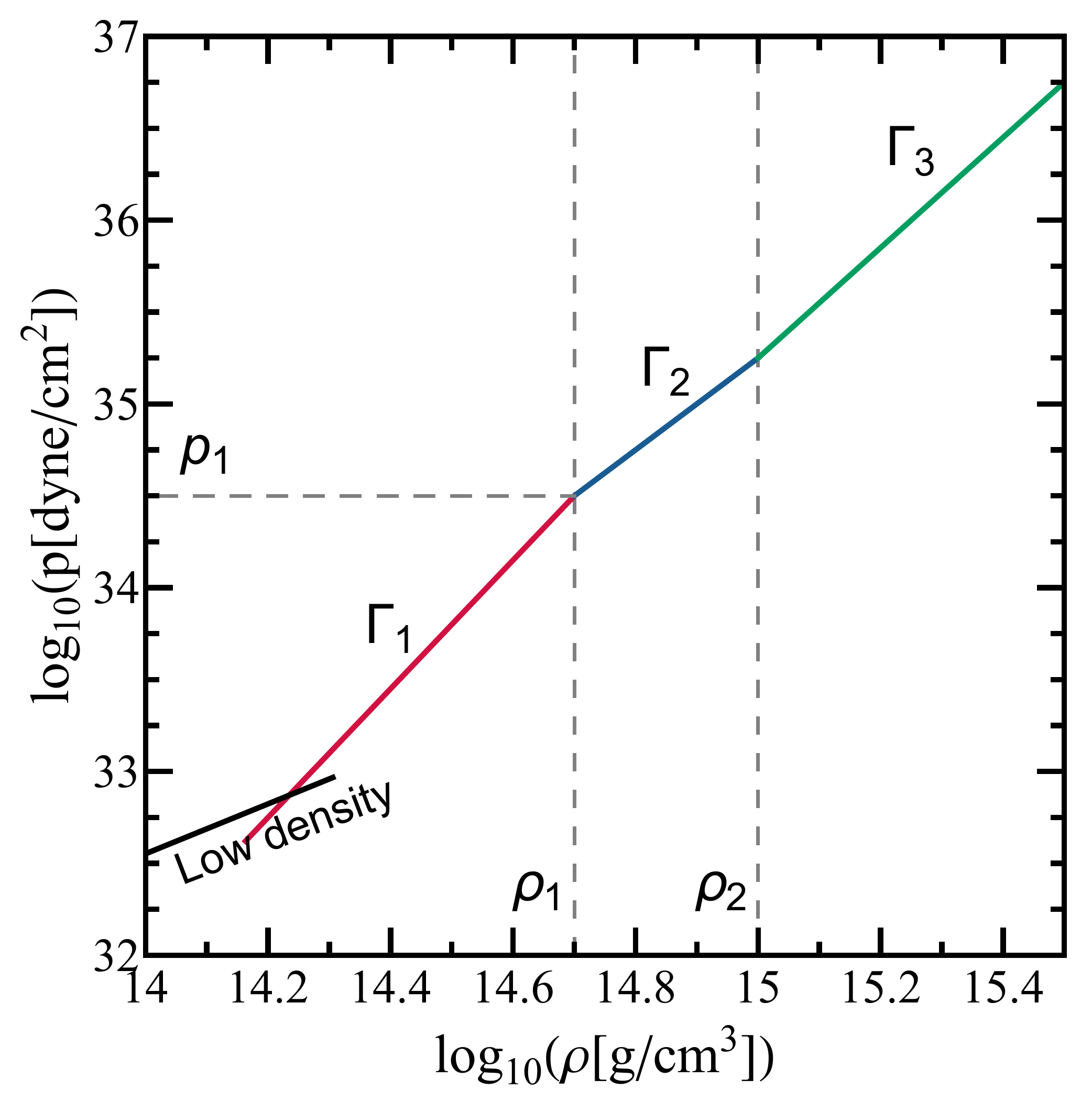}
\caption{Schematic representation of the dividing regions which
specify the piecewise representation of the NS structure. The inner
and center pieces are separated by the dividing densities
$\rho_{1,2}$, fixed to $\rho_1 = \text{10}^{\text{14.7}} \,
\text{g/cm}^{\text{3}}$  and $\rho_2 = \text{10}^{\text{15}} \,
\text{g/cm}^{\text{3}}$, respectively. Adapted from \cite{read2009}.}
\label{fig:piecewise} \end{figure}

Piecewise polytropes accurately fit the energy-density profiles of a
large variety of EoSs based on realistic nuclear-physics calculations.
These include pure nucleonic matter, hyperons, meson condensates and
phase transitions to deconfined quarks. The NS macroscopic
observables, like masses and radii, are accurately reproduced within
$\lesssim 1\%$ of the corresponding ``exact'' values.
This accuracy is achieved requiring that the high-density core is represented by three polytropic segments
\begin{equation}
p(\rho) = K_i \rho^{\Gamma_i}   \qquad \rho_{i-1} \leq \rho \leq \rho_{i}, 
\label{eq:polytropic}
\end{equation}
specified by the dividing rest-mass densities, with adiabatic constant
and index given by  $K_i$ and $\Gamma_i$, respectively. Read and
collaborators found that the values of the dividing densities which
minimize the discrepancy with respect to the tabulated EoSs correspond
to  $\rho_1 = \text{10}^{\text{14.7}} \, \text{g/cm}^{\text{3}}$  and
$\rho_2 = \text{10}^{\text{15}} \, \text{g/cm}^{\text{3}}$. A
schematic picture of this model is shown in Fig.~\ref{fig:piecewise}.

The energy density $\epsilon$ is given by the integral of the first law of thermodynamics:
\begin{equation}
d\epsilon = \frac{(\epsilon+p)}{\rho} d\rho\ , 
\end{equation}
which can be recast through Eq.~(\ref{eq:polytropic}) to the following form 
\begin{equation}
\epsilon(\rho) = (1+ a_i )\rho + \frac{p(\rho)}{\Gamma_i-1},
\end{equation}
where $a_i$ are integration constants to be determined requiring
continuity between the three regions.  Specifying the initial density
of the outer interface reduces the number of independent parameters to
four variables, namely $\{p_1,\Gamma_1,\Gamma_2,\Gamma_3\}$, where $p_1 \equiv \log_{10} p(\rho_1)$.
As we will discuss in the next section, the latter is the actual
quantity to be constrained through the inverse stellar approach. 

At low densities, the last interface is matched dynamically to a fixed
crust, which is chosen to be a parametrized four-piece polytropic
version of the SLy EoS. The matching point is simply given by the
value of density where the crust and core-EoS intersect each other,
and depends only on $p_1$ and $\Gamma_1$. This choice naturally
implies a constraint on these two parameters, since specific
combinations of $p_1$ and $ \Gamma_1$ do exist, which yield no
intersection between the crust and the core EoSs and are therefore
incompatible.  The allowed region can be found analytically, and
satisfies the following relation \begin{equation}\label{constraint}
\Gamma_1 > \frac{\log{\left(
\frac{p(\rho_1)}{p_\tn{sly}}\right)}}{\log{\left(
\frac{\rho_1}{\rho_\tn{sly}}\right)}}, \end{equation} where
$\rho_\tn{sly}$ and $p_\tn{sly}$ are the density and the pressure in
the inner region of the crust.

\begin{figure}[th]
\centering
\includegraphics[width=4.25cm]{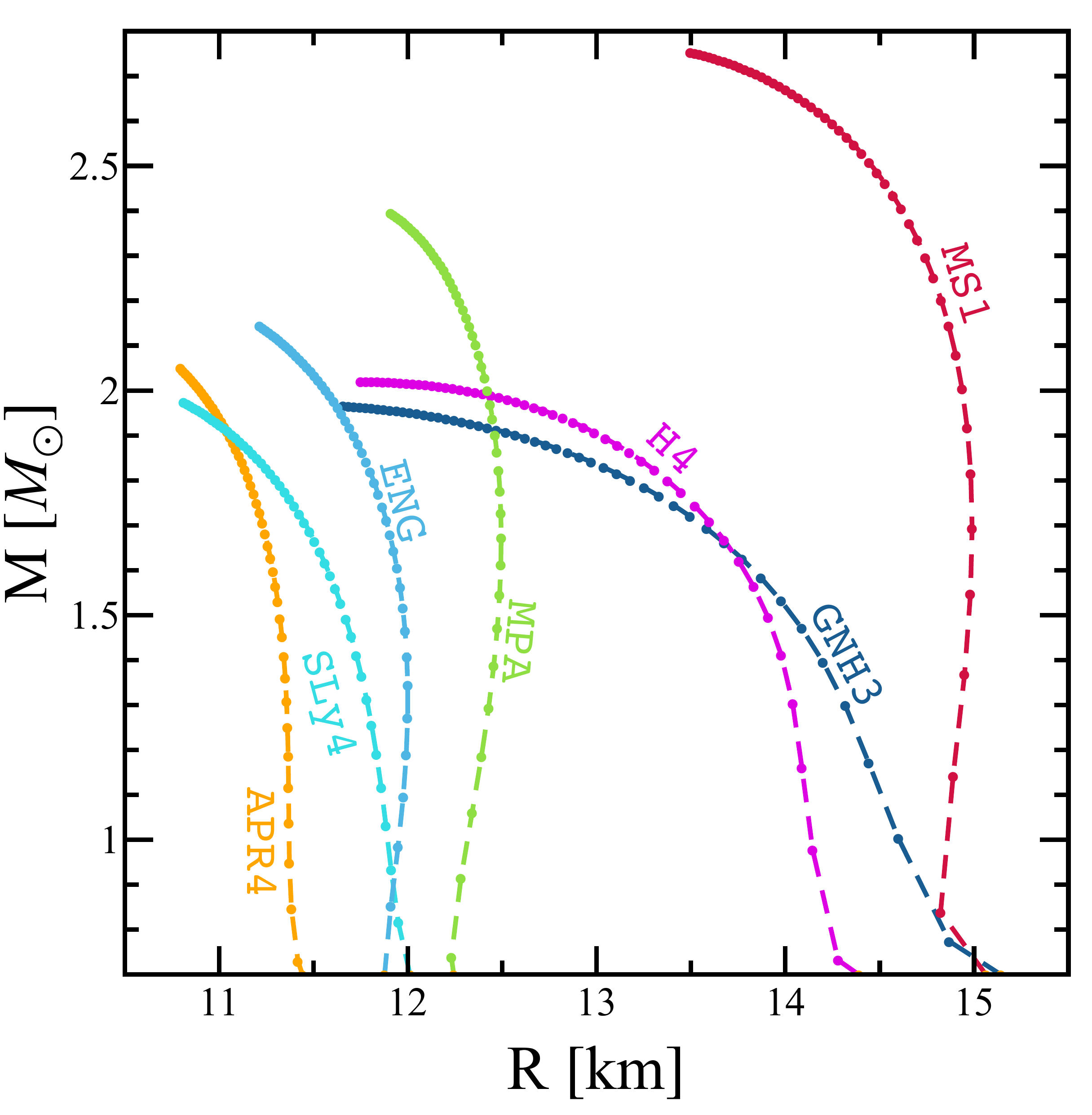}
\includegraphics[width=4.25cm]{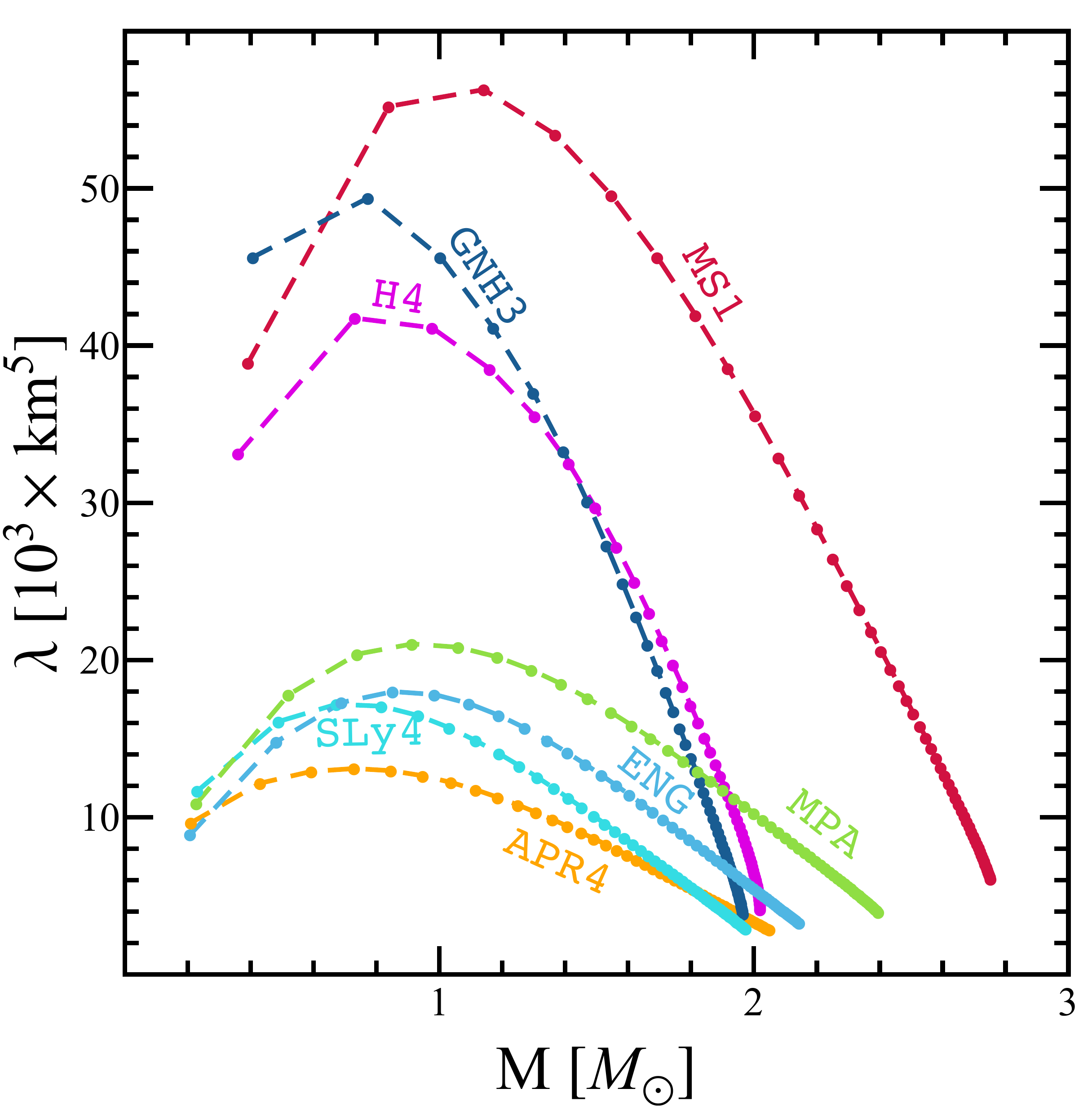}
\caption{(Left) Mass-radius relations for some realistic EoSs modeled
through piecewise polytropes.  The values of the parameters
$\{p_1,\Gamma_1,\Gamma_2,\Gamma_3\}$ which specify the EoS can be
found in Table III of \cite{read2009}. (Right) Tidal deformability
$\lambda$ as a function of the NS mass for the same EoS considered in
the left panel.} \label{fig:LoveMass} \end{figure}

Figure~\ref{fig:LoveMass} shows the mass-radius relations obtained
by solving the relativistic equations of stellar structure for
different EoSs, modeled through the piecewise parametrization. Details
on the various EoS can be found in Section II of \cite{read2009}. 

\section{The Markov Chain Monte Carlo}\label{Sec:MCMC}

In this section we shall describe the numerical approach we use to
estimate the piecewise parameters, starting from the macroscopic
observables provided by gravitational wave observations, namely the
mass $M$ and the tidal deformability $\lambda$ of detected NSs. It is
worth noting that our method is completely general, and can be
applied also using different NS observables, obtained either with electromagnetic or
gravitational wave observations, leading to a genuine multimessenger
framework \cite{paper2}.

In general, for a given set of $N$ observed stars, we have $m+N$ free
parameters to determine, i.e., $m$ parameters of the EoS
model, and $N$ central pressures $p^c_{i=1\ldots N}$.
Any detected NS provides $2$ observables, which we have assumed to be
the mass and the tidal deformability. Therefore, to fully characterize the
parametrized EoS, we need at least $N=m$ observations. 

As discussed in
Sec.~\ref{Sec:EOS},  piecewise polytropes are characterized by $m=4$
parameters, which lead to $8$ unknown parameters to be found: 
\[
\vec{\theta}=\{p_1,\Gamma_1,\Gamma_2,\Gamma_3,p^c_1,p^c_2,p^c_3,p^c_4\};
\]
therefore, we need  at least  $4$ observations, which provide the required set of
$8$ measured quantities
$\vec{d}=\{M_i,\lambda_i\}_{i=1\ldots4}$.

Within Bayesian inference, we are interested in determining the
posterior probability density function (PDF) of the EoS parameters
given the experimental data: $\mathcal{P} (\vec{\theta}\vert\vec{d}
)$. Using Bayes' theorem, we can write the joint PDF for EoS parameters
as: 
\begin{equation} \mathcal{P} (\vec{\theta}\vert\vec{d})  \propto
\mathcal{P}_0(\vec{\theta}) \mathcal{L} (\vec{d}\vert\vec{\theta}) \ ,
\label{eq:PDF} 
\end{equation} 
where $\mathcal{P}_0(\vec{\theta})$
describes the prior information on the parameters, and
$\mathcal{L}(\vec{d}\vert\vec{\theta})$ is the likelihood function.
The probability distribution of $l$ parameters is given by
marginalizing over the remaining $N+4-l$ variables, i.e.,
\begin{equation} {\cal P}(\theta_1,\ldots,\theta_{l})=\int {\cal
P}(\vec{\theta})d\theta_{l+1}\ldots d\theta_{N+4}\ .  \end{equation}
In our analysis we assume that the set of data $\vec{d}$
obtained from GW detections are independent and Gaussian distributed,
with the values of each observable $M_i\ (\lambda_i)$ being affected
by an experimental uncertainty $\sigma_{M_i}\ (\sigma_{\lambda_i})$.
Under this assumptions the likelihood can be written as ${\cal
L}\propto\mathrm{e}^{-\chi^2}$, where the chi-square variable reads:
\begin{align} \chi^2 = \frac{1}{2}\sum_{i=1}^N&\left\{\frac{\left[ M
\left(p_1,\Gamma_1,\Gamma_2,\Gamma_3,p^c_i \right)-M_i
\right]^2}{\sigma_{M_i}^2} \right.\nonumber\\
&\phantom{aaaaaa}\left.+\frac{\left[\lambda
\left(p_1,\Gamma_1,\Gamma_2,\Gamma_3,p^c_i \right)-\lambda_i
\right]^2}{\sigma_{\lambda_i}^2}\right\}.  \end{align}

We sample the posterior probability distribution (\ref{eq:PDF}) using
Markov chain Monte Carlo (MCMC) simulations based on the
Metropolis-Hastings algorithm \cite{gilks1995markov}. The procedure of
this framework can be summarized with the following steps.

Given an initial point $\vec{\theta}_1=\left\{p_1,\Gamma_1,\Gamma_2,\Gamma_3,p^c_{i=1\ldots N}\right\}$, randomly chosen 
within the parameter space, we propose a jump to a new state,  $\vec{\theta}_2$, with probability specified by the proposal 
function $f=f(\vec{\theta}_1,\vec{\theta}_2)$. The latter is chosen to be a multivariate Gaussian distribution\footnote{Note that with this choice $f$ is symmetric, i.e., $f(\vec{\theta}_2,\vec{\theta}_1)=f(\vec{\theta}_1,\vec{\theta}_2)$.} centered 
in the current state $\vec{\theta}_1$, $f(\vec{\theta}_2,\vec{\theta}_1) = \mathcal{N}\left(\vec{\theta}_1, \mathbf{\Sigma} \right)$.
Then, we compute the ratio
\begin{equation}
r(\vec{\theta}_1,\vec{\theta}_2) = \frac{\mathcal{P}(\vec{\theta}_2)}{\mathcal{P}(\vec{\theta}_1)}\ ,
\end{equation}
and accept the proposed move with probability
\begin{equation}
a(\vec{\theta}_1,\vec{\theta}_2)=\mathrm{min}\left\{1,r(\vec{\theta}_1,\vec{\theta}_2)\right\}.
\end{equation}
In this way, the chain is updated to the step $\vec{\theta}_2$ with probability $a(\vec{\theta}_1,\vec{\theta}_2)$, or 
remains fixed in $\vec{\theta}_1$ with probability $1-a(\vec{\theta}_1,\vec{\theta}_2)$. If $\mathcal{P}(\vec{\theta}_2) 
\geq \mathcal{P}(\vec{\theta}_1)$ the jump is always accepted, while if $\mathcal{P}(\vec{\theta}_2) < \mathcal{P}(\vec{\theta}_1)$ 
it is accepted with probability $r(\vec{\theta}_1,\vec{\theta}_2)$. The previous steps are then iterated $n$ times, 
allowing the chain to explore the parameter space of the model (a workflow is shown in Algorithm \ref{mh}). 
MCMC theory guarantees that, from any initial state and proposal function, the system evolves towards the desired 
target distribution $\mathcal{P}(\vec{\theta})$. In practical situations however, the convergence of the chain is strongly 
affected by the choice of the proposal function. In this paper we adopt an adaptive framework, in which the 
covariance $\mathbf{\Sigma}$ of $f(\vec{\theta}_1,\vec{\theta}_2)$ is continuously updated though a {\it Gaussian adaptation} algorithm (GaA) \cite{5586491}.  A remarkable feature of this approach is that the acceptance probability $P$ of the 
proposed jump can be fixed a priori (a detailed description of the formalism is presented in Appendix~\ref{Sec:appA}).

\section{Numerical setup}\label{Sec:setup}

In order to test the ability of our approach to reconstruct the
parameters of the piecewise polytropes, we have analyzed different
possible scenarios. We consider nonspinning NSs with
$M\in[1.1-1.6]M_\odot$, which covers most of the mass range determined
so far by electromagnetic observations of binary pulsars
\cite{Ozel:2016oaf}.  Moreover, we focus on two EoS, \texttt{apr4} and
\texttt{h4}. As shown in Fig.~\ref{fig:LoveMass}, these models span a
wide range of mass-radius configurations. Moreover they fit within the
$90\%$ credible interval estimated by the LIGO/Virgo collaboration
after the first GW detection from a binary NS
\cite{PhysRevLett.119.161101, 2041-8205-848-2-L13}. Therefore,
\texttt{apr4} and \texttt{h4} are the best candidates to represent
extreme cases of {\it soft} and {\it stiff} nuclear matter, compatible
with astrophysical observations.  For both EoSs, we  compare the
features of a canonical $1.4M_\odot$ NS in
Table~\ref{table:massradius}, which also shows how the tidal
deformability of the two EoS differs by a factor $>3$. We remember
that large values of $\lambda$ yield stronger changes in the GW
signal, and therefore lead to tighter constraints.
 
\newcolumntype{Y}{>{\centering\arraybackslash}X}
\begin{table}[ht]	
  \centering
    \caption{Radius and tidal deformability of prototype $1.4M_\odot$ NSs modeled with \texttt{apr4} and \texttt{h4}. }
  \begin{tabularx}{8.5cm}{YYY}
     \hline
    \hline
\texttt{EoS} &  $R_\tn{NS} $ [km] & $\lambda$ [km$^5$]\\
     \hline
\texttt{apr4} &  11.34  & 9502 \\ 
\texttt{h4} & 13.99 & 32861 \\
   \hline
  \hline
\end{tabularx}
  \label{table:massradius}
\end{table}
It is important to stress that for $M\lesssim 1.6M_\odot$ the
adiabatic index $\Gamma_3$ does not affect the structure of the star
for both \texttt{apr4} and \texttt{h4}. Therefore, we can safely
neglect this coefficient within the analysis, reducing 
the parameter space volume to
$\vec{\theta}=\{p_1,\Gamma_1,\Gamma_2,p^c_1,p^c_2,p^c_3\}$. Note that
the EoS is now fully specified by only three variables, and as a
consequence we only need six observables, which correspond to three
observed NSs. 
No observational constraints have been derived so far for the Love
numbers of a neutron star \footnote{The gravitational wave event
GW170817 has actually allowed to set an upper bound on the {\it
average} tidal deformability of the two stars, $\tilde{\Lambda}\leq
800$, where
$$\tilde{\Lambda}=\frac{16}{13m^5}\left[\frac{m_1+12m_2}{m_1}\lambda_1+\frac{m_2+12m_1}{m_2}\lambda_2\right]\
,$$ being $(m_1+m_2)$ the total mass of the binary
\cite{PhysRevLett.119.161101}.}.  Therefore, we compute
$\sigma_\lambda$ using a Fisher matrix approach
\cite{Vallisneri:2007ev}, and considering equal-mass binary NSs at
prototype distance of $100$ Mpc. 

The uncertainties are computed for
the advanced generation of detectors.  More specifically, we will
assume that the GW events have been detected by a
network\footnote{Note that for a $N$ independent interferometers the
error on the Love number is roughly reduced by a factor
$\sim1/\sqrt{N}$, with respect to the single detector analysis.} of
interferometers (HLVK)  composed by the two LIGO sites, Virgo and the
Japanese KAGRA. For all the measurements we consider detector's
configurations at design sensitivity \cite{zerodet,virgo,kagra}.
Moreover, following \cite{0004-637X-784-2-119}, we fix the uncertainty
on the NS mass, $\sigma_M$ to $10\%$ of the measured value for HLVK.
We choose flat prior distributions for all the parameters of the
piecewise EoS, within the range $p_1\in[33,35]$,
$\Gamma_{1,2}\in [1,4]$ and $p^c_i\in[10^{-6},10^{-3}]\ \tn{km}^{-2}$.
The adiabatic index and the initial pressure of the outer core,
$(\Gamma_1,p_1)$, are also constrained by the theoretical bound given
by Eq.~\eqref{constraint}. Finally, for each set of data, we run four
parallel processes of $n=5\times 10^5$ samples.  We assess the
convergence of the MCMC simulations to the target distribution through
the Rubin test, and by analyzing the autocorrelation of each chain
\cite{gilks1995markov}. 

\begin{figure*}
\centering
\includegraphics[width=4.5cm]{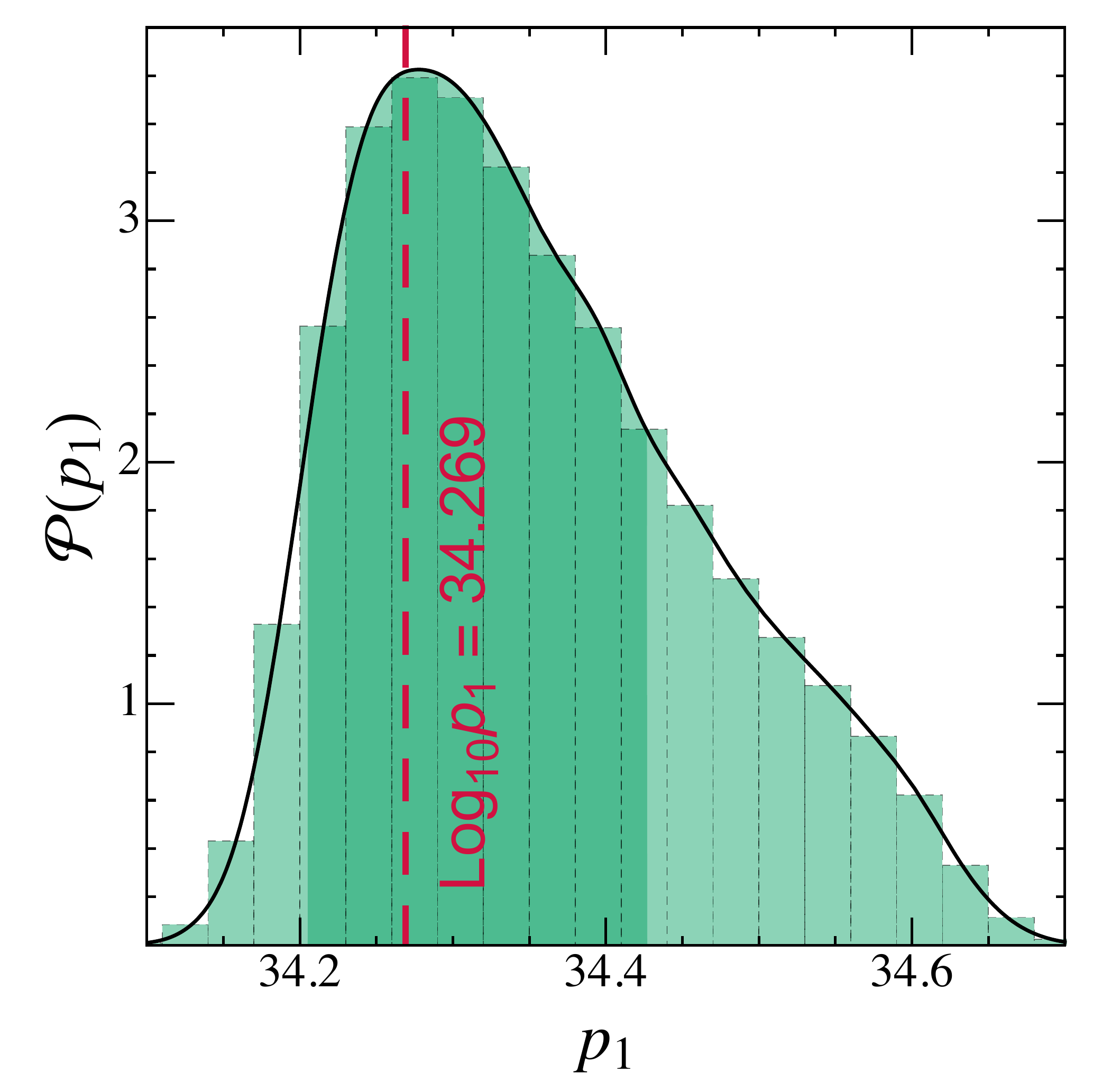}
\includegraphics[width=4.5cm]{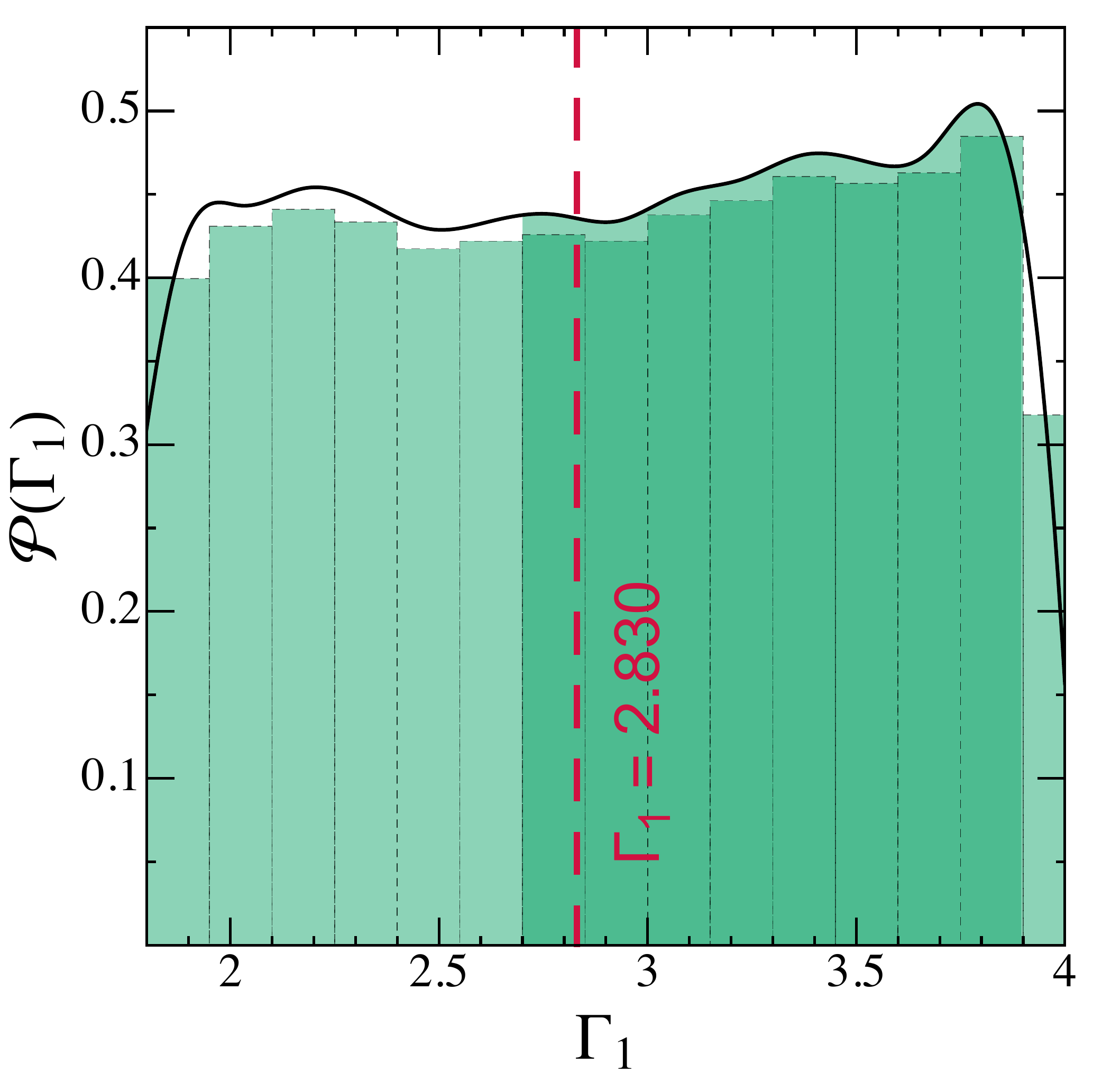}
\includegraphics[width=4.5cm]{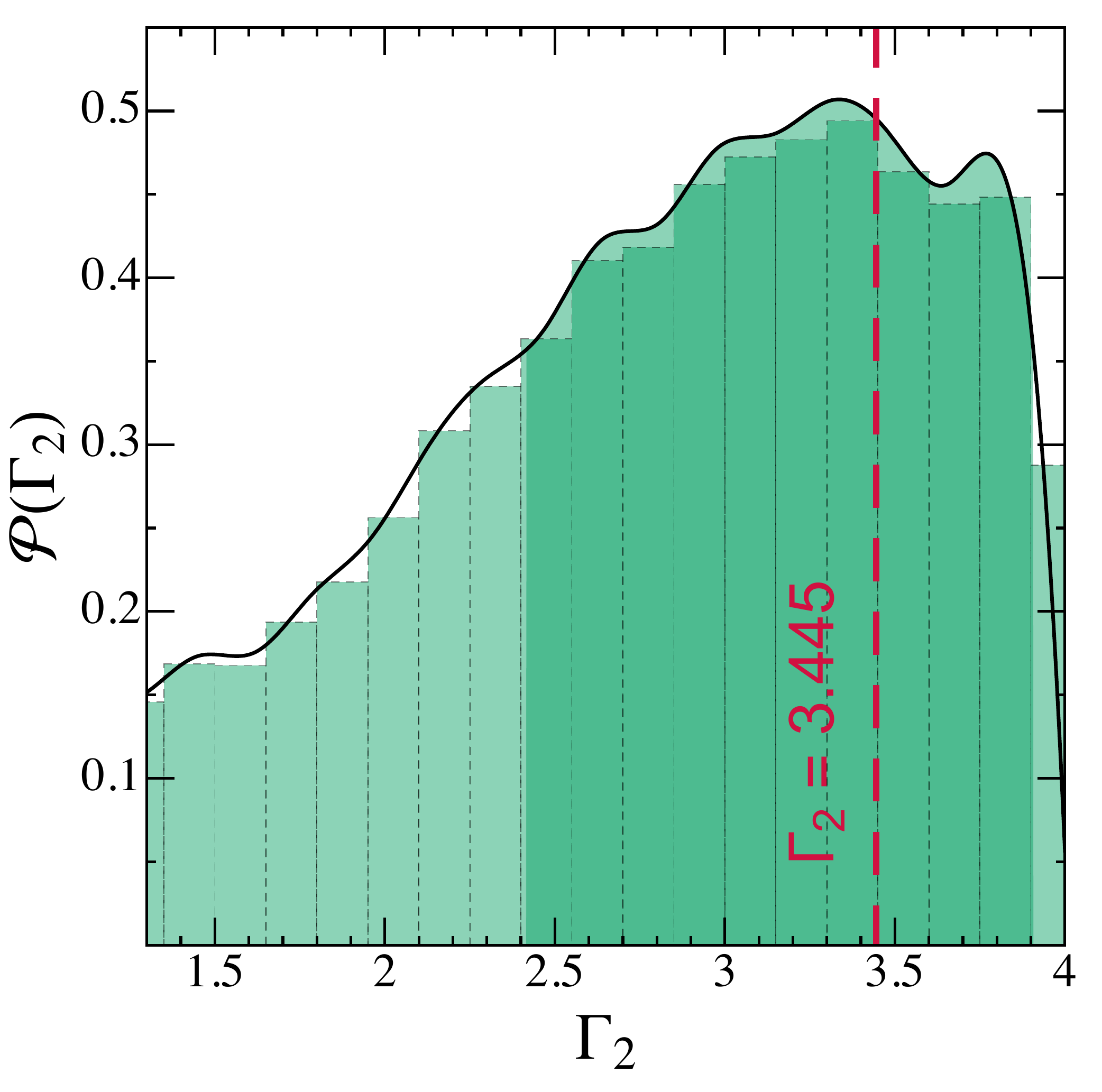}\\
\includegraphics[width=4.5cm]{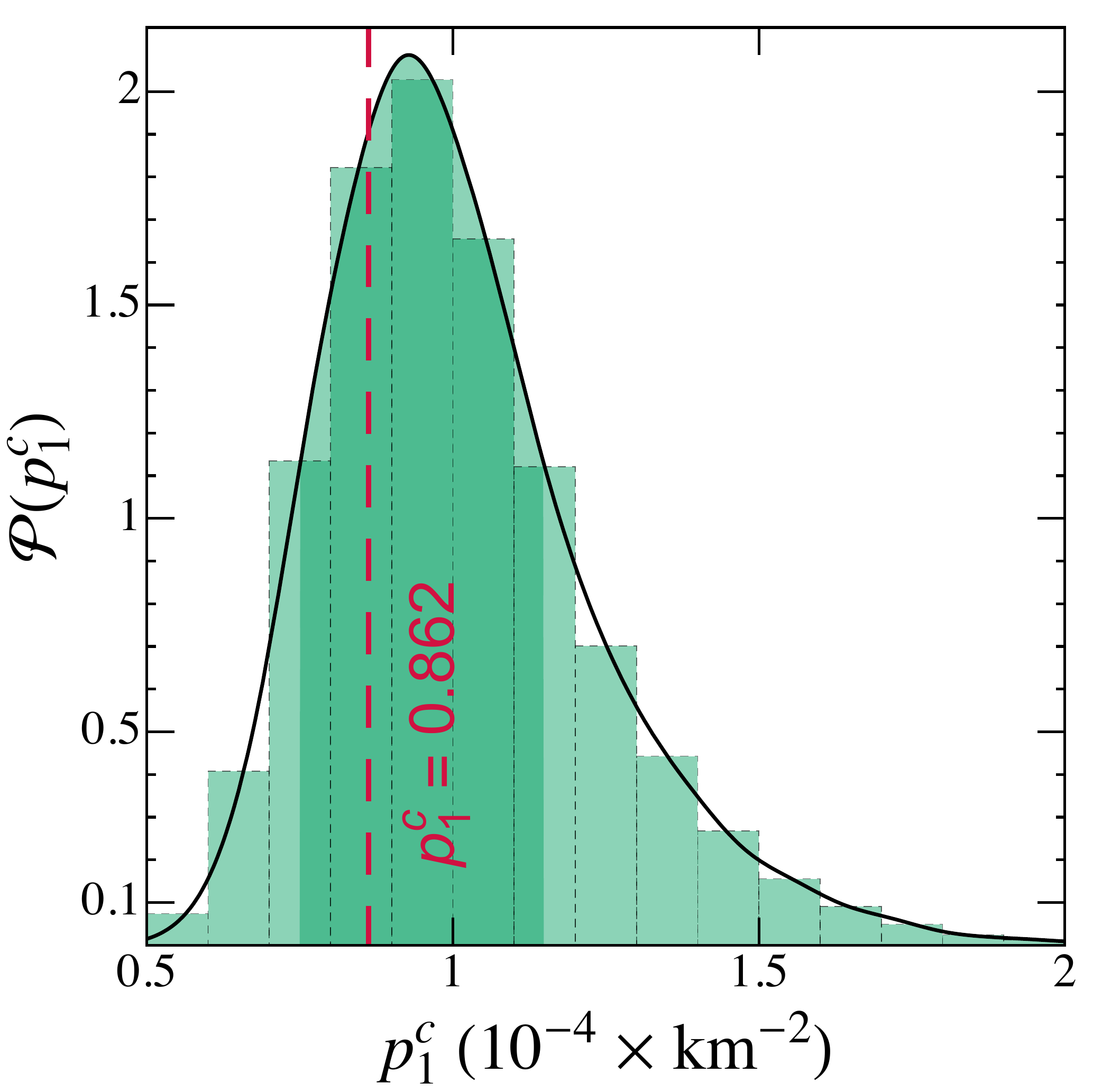}
\includegraphics[width=4.5cm]{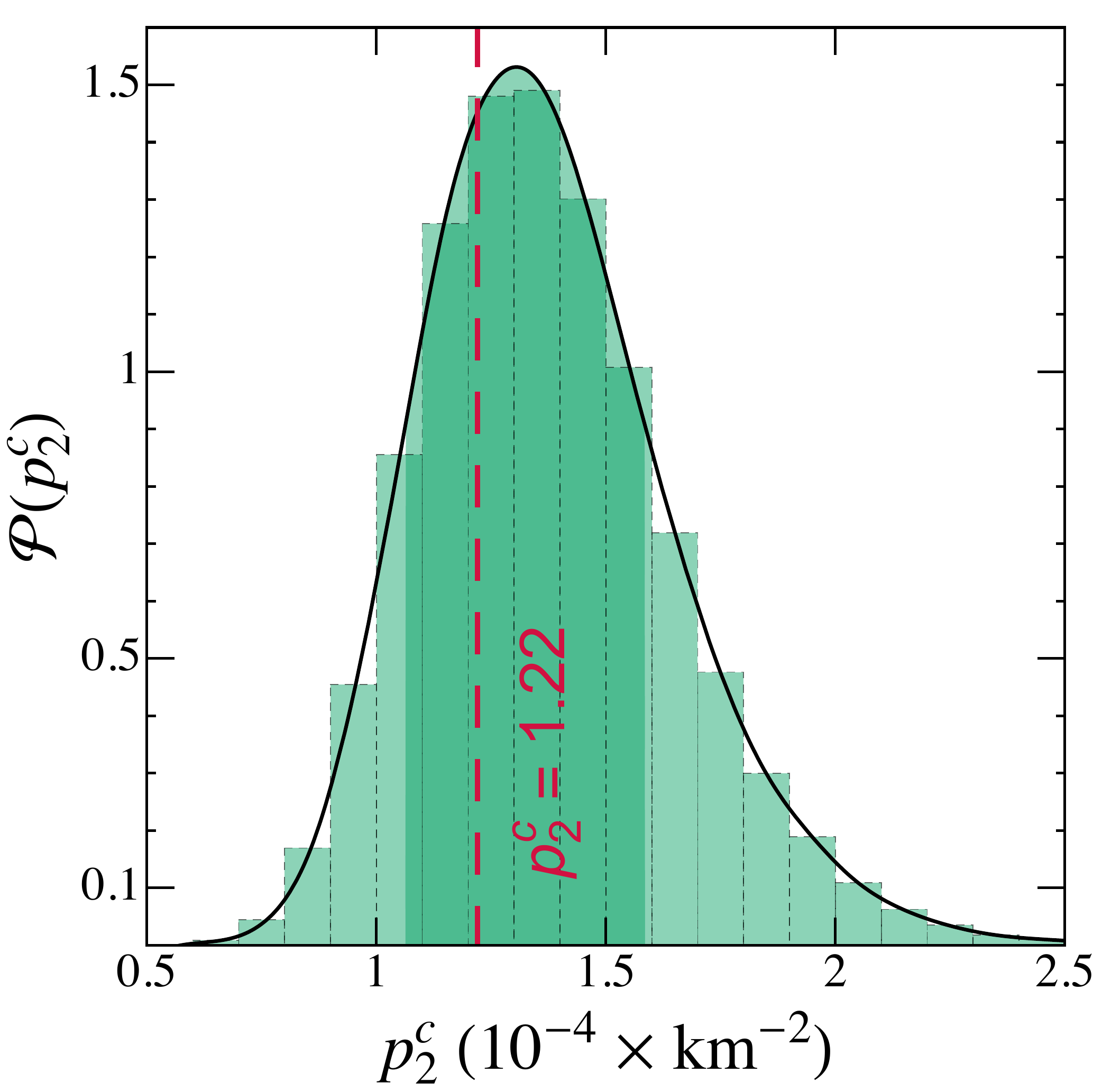}
\includegraphics[width=4.5cm]{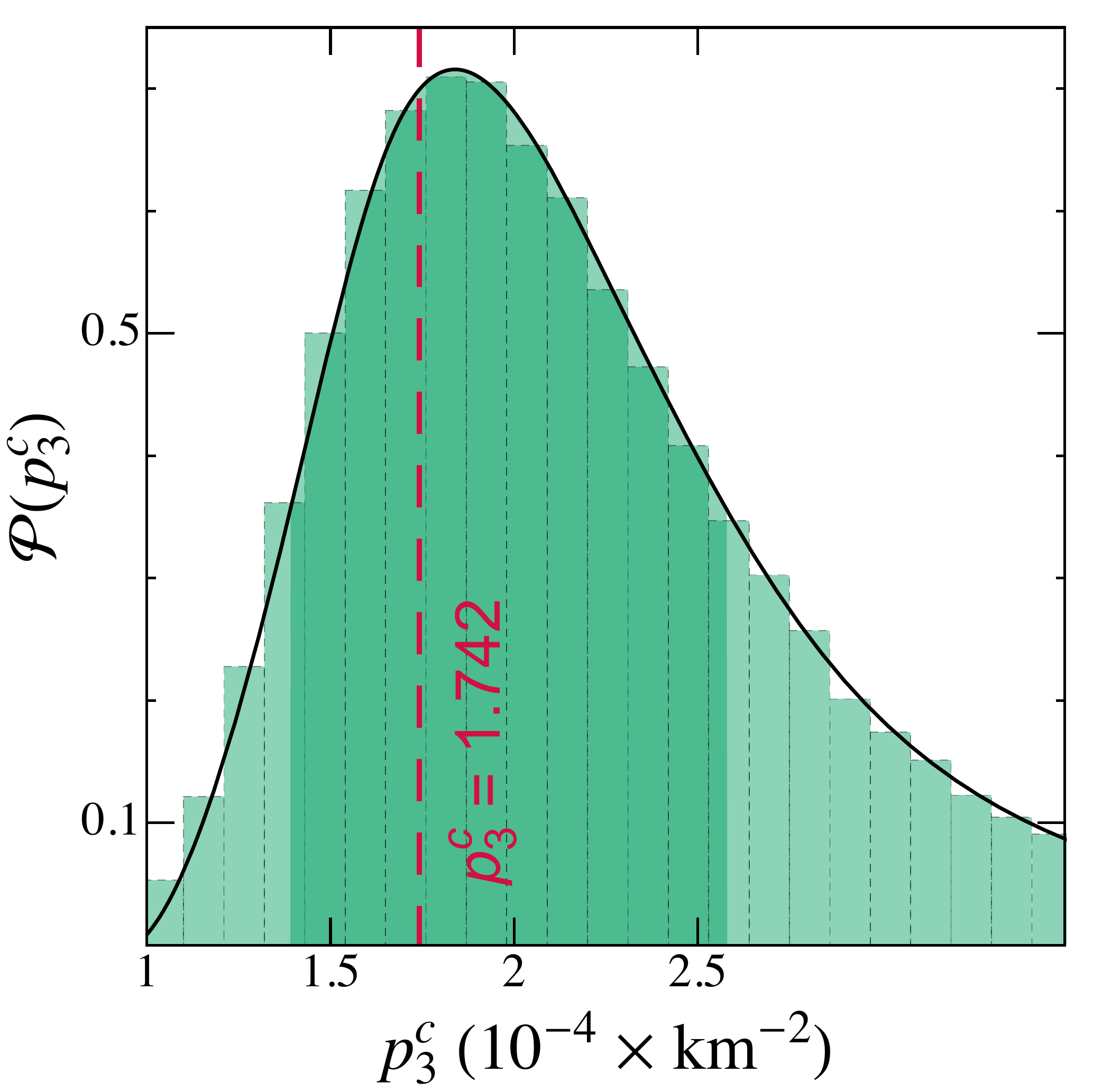}
\caption{Marginalized probability distributions for the piecewise
parameters of the \texttt{apr4} EoS, derived for the \texttt{m246} model
with NS masses $(1.2,1.4,1.6)M_\odot$, obtained for the network HLVK.
The histograms of the sampled points are showed below each function.
Dashed vertical lines identify the injected true values, while the
shaded bands correspond to the $1\sigma$ credible regions of each
parameter.} \label{fig:Love} \end{figure*}

\begin{table*}[ht]	
  \centering
    \caption{Comparison between injected and reconstructed values of the
\texttt{apr4} and \texttt{h4} parameters for the three models 
  analyzed in this paper. For each parameter of the piecewise EoS we show the $1\sigma$ credible level of the marginalized distribution.}
  \begin{tabular}{ccccc|ccc|ccc}
& &  & \texttt{m246} & & & \texttt{m456} & & & \texttt{m123}\\
EoS&  & injected & & $1\sigma$ &  injected & & $1\sigma$ & injected & & $1\sigma$\\
     \hline
\texttt{apr4}&$p_1$ & 34.269 & & [34.205 - 34.427] &  34.269 && [34.247 - 34.582]  & 34.269 && [34.209 - 34.367]\\ 
&$\Gamma_1$ &  2.830  & &  [2.700 - 3.896]  & 2.830 &&  [2.212 - 3.846] & 2.830 && [2.458 - 3.898]\\
&$\Gamma_2$ & 3.445 & & [2.415 - 3.907] & 3.445 && [1.817 - 3.599] & 3.445 && [2.691 - 3.952]\\
&$10^{-4}\times p_1^c $ & 0.862 & & [0.750 - 1.15] & 1.22 && [1.09 - 1.76] &  0.722 &&  [0.623 - 0.919]\\
&$10^{-4}\times p_2^c$ & 1.22 & & [1.06 - 1.58] & 1.45 && [1.29 - 2.12] &  0.862 &&  [0.752 - 1.07]\\
&$10^{-4}\times p_3^c$ & 1.74 & & [1.39 - 2.58] & 1.74 && [1.46 - 2.70] &   1.03 &&  [0.893 - 1.26]\\
\hline
\texttt{h4}&$p_1$ & 34.669 & & [34.611 - 34.738] &  34.669 && [34.628 - 34.742]  & 34.669 && [34.644 - 34.771]\\ 
&$\Gamma_1$ &  2.909  & &  [2.479 - 3.401]  & 2.909 &&  [1.956 - 3.906] & 2.909 && [2.752 - 3.520]\\
&$\Gamma_2$ & 2.246 & & [1.732 - 3.518] & 2.246 && [1.056 - 2.383] & 2.246 && [1.055 - 3.596]\\
&$10^{-4}\times p_1^c$ & 0.372 & & [0.310 - 0.446] & 0.533 && [0.423 - 0.643] &  0.311 &&  [0.260 - 0.355]\\
&$10^{-4}\times p_2^c$ & 0.533 & & [0.486 - 0.614] & 0.650 && [0.556 - 0.773] &  0.372 &&  [0.330 - 0.427]\\
&$10^{-4}\times p_3^c$ & 0.804 & &  [0.721 - 0.930] & 0.804 && [0.706 - 0.957] &   0.443 &&  [0.407 - 0.512]\\
   \hline
  \hline
\end{tabular}
  \label{table:injected}
\end{table*}

\begin{figure}[th]
\centering
\includegraphics[width=4.2cm]{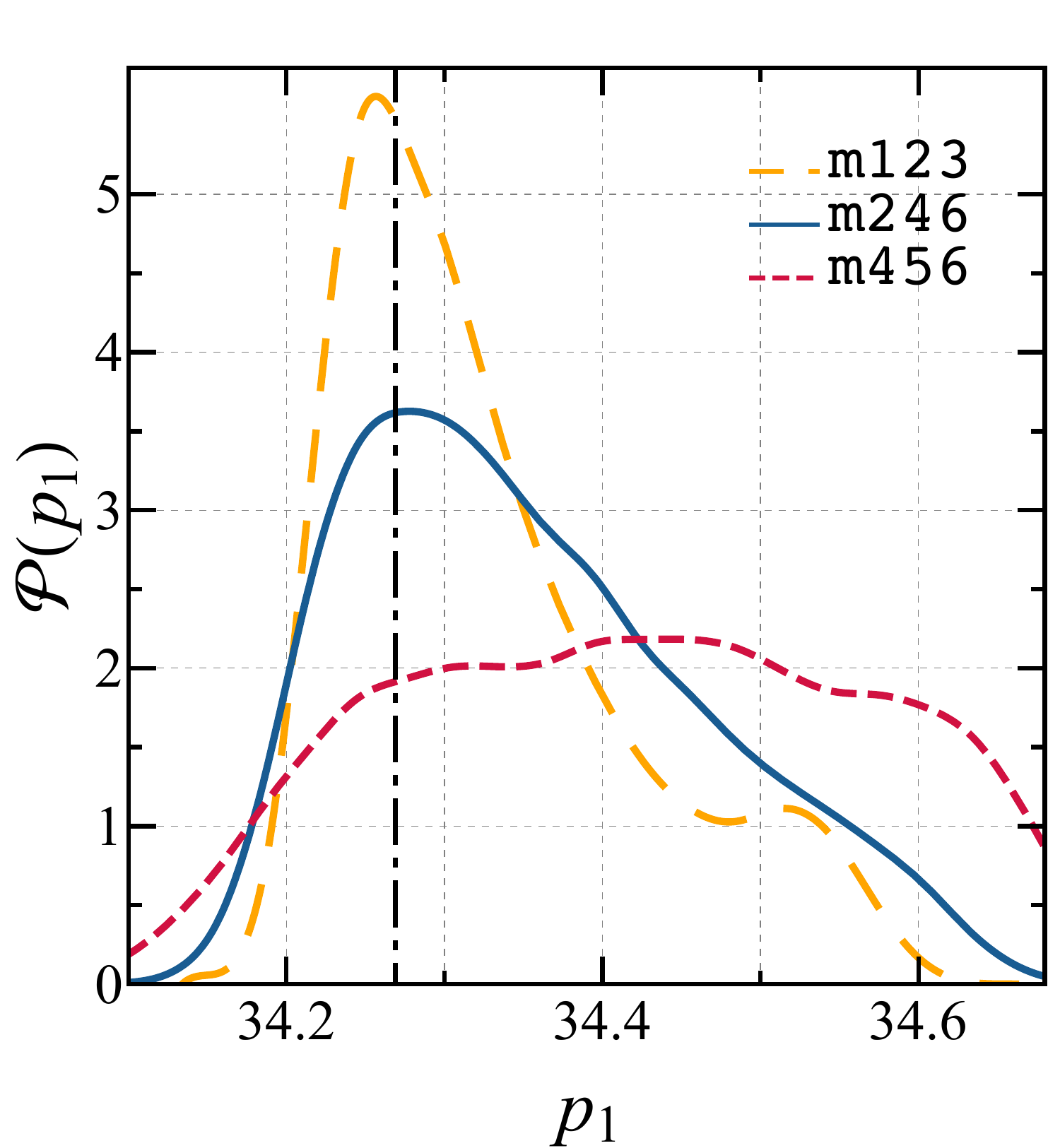}
\includegraphics[width=4.35cm]{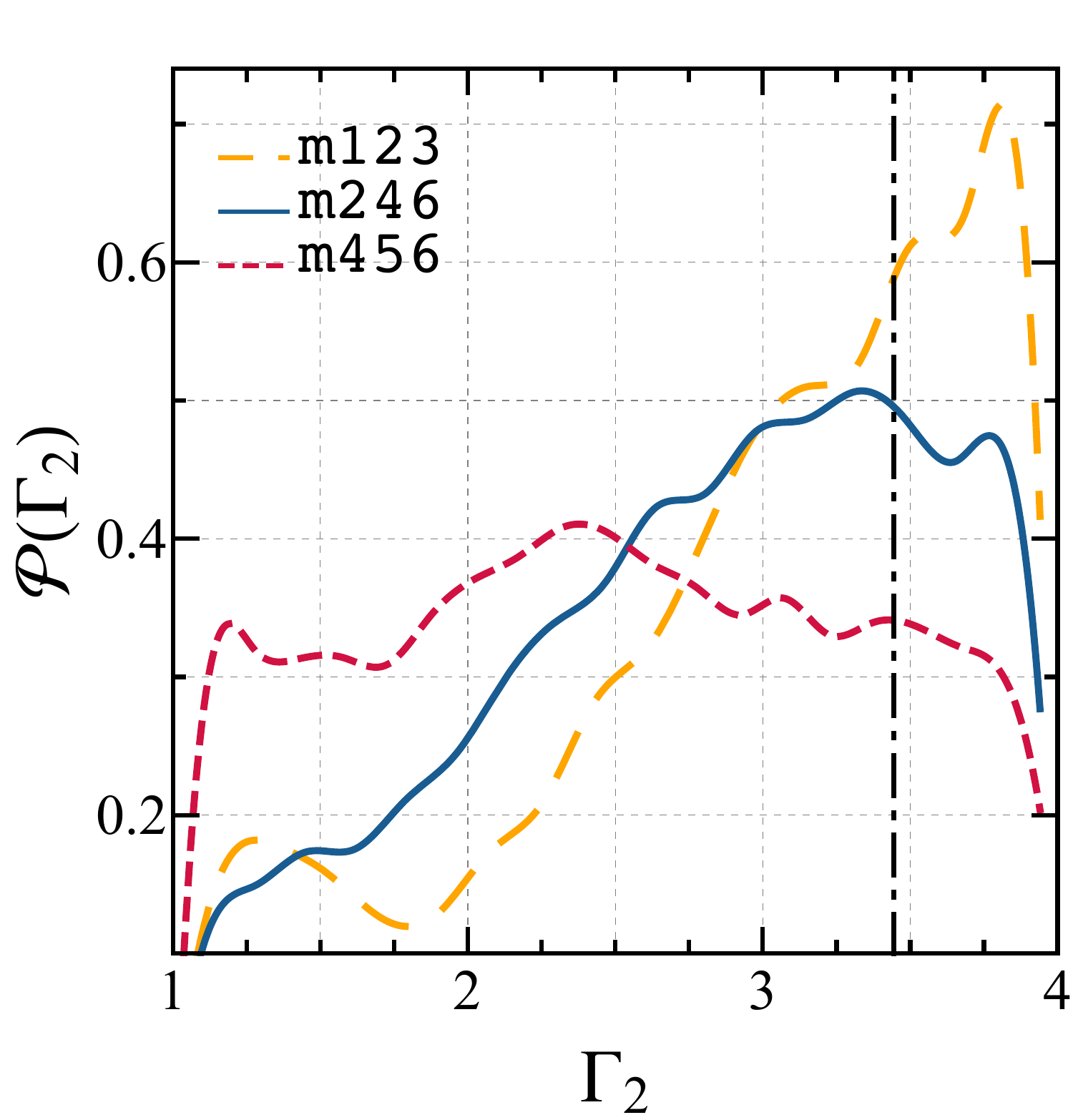}
\caption{Comparison among the marginalized posterior of $p_1$ (left)
and $\Gamma_2$ (right) for \texttt{apr4},
derived for the models \texttt{m246}, \texttt{m456} and \texttt{m123}. The dot-dashed
vertical lines correspond to the true values of the parameters.}
\label{fig:Love3}
\end{figure}
\begin{figure*}[th]
\centering
\includegraphics[width=5.5cm]{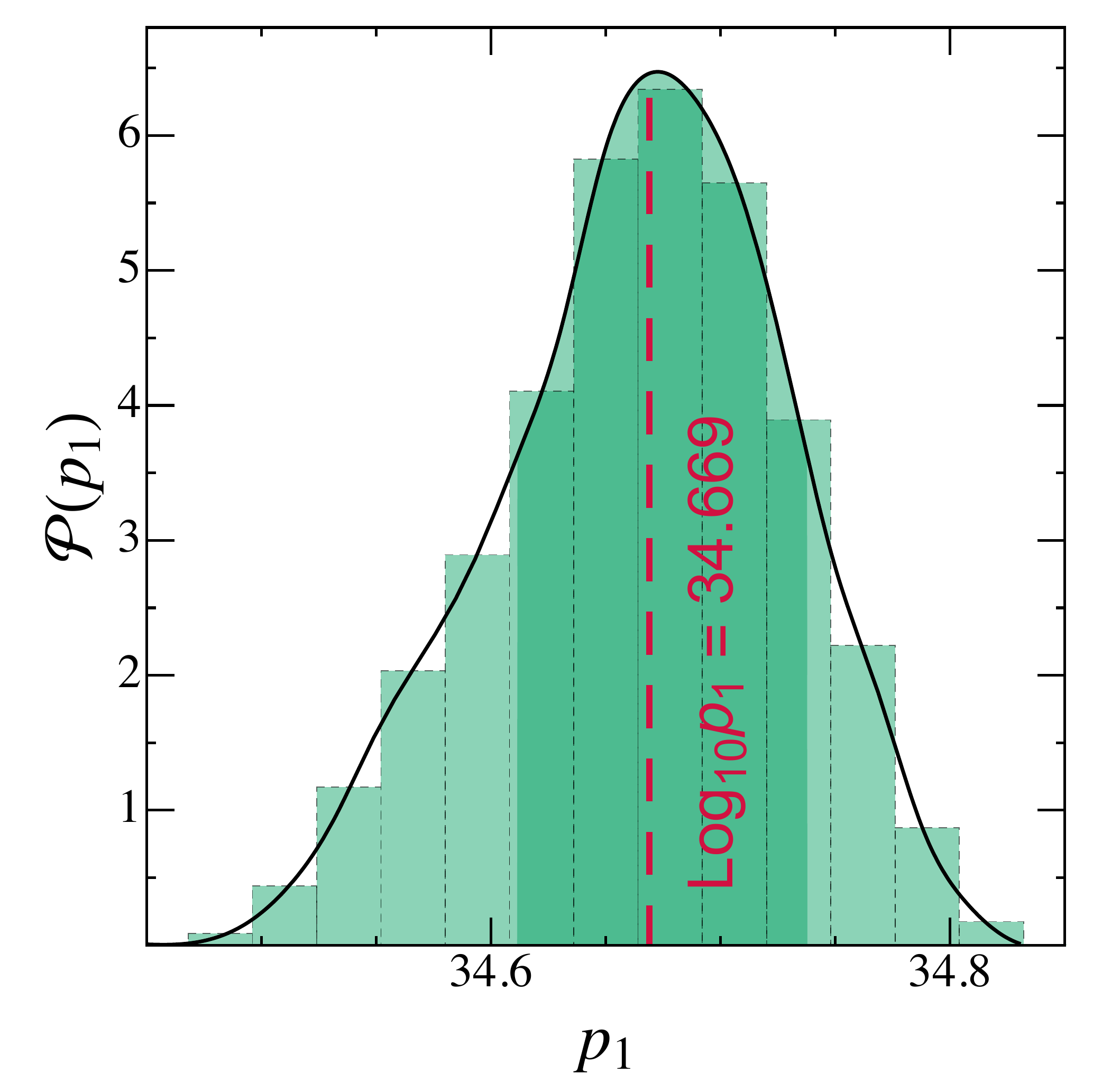}
\includegraphics[width=5.5cm]{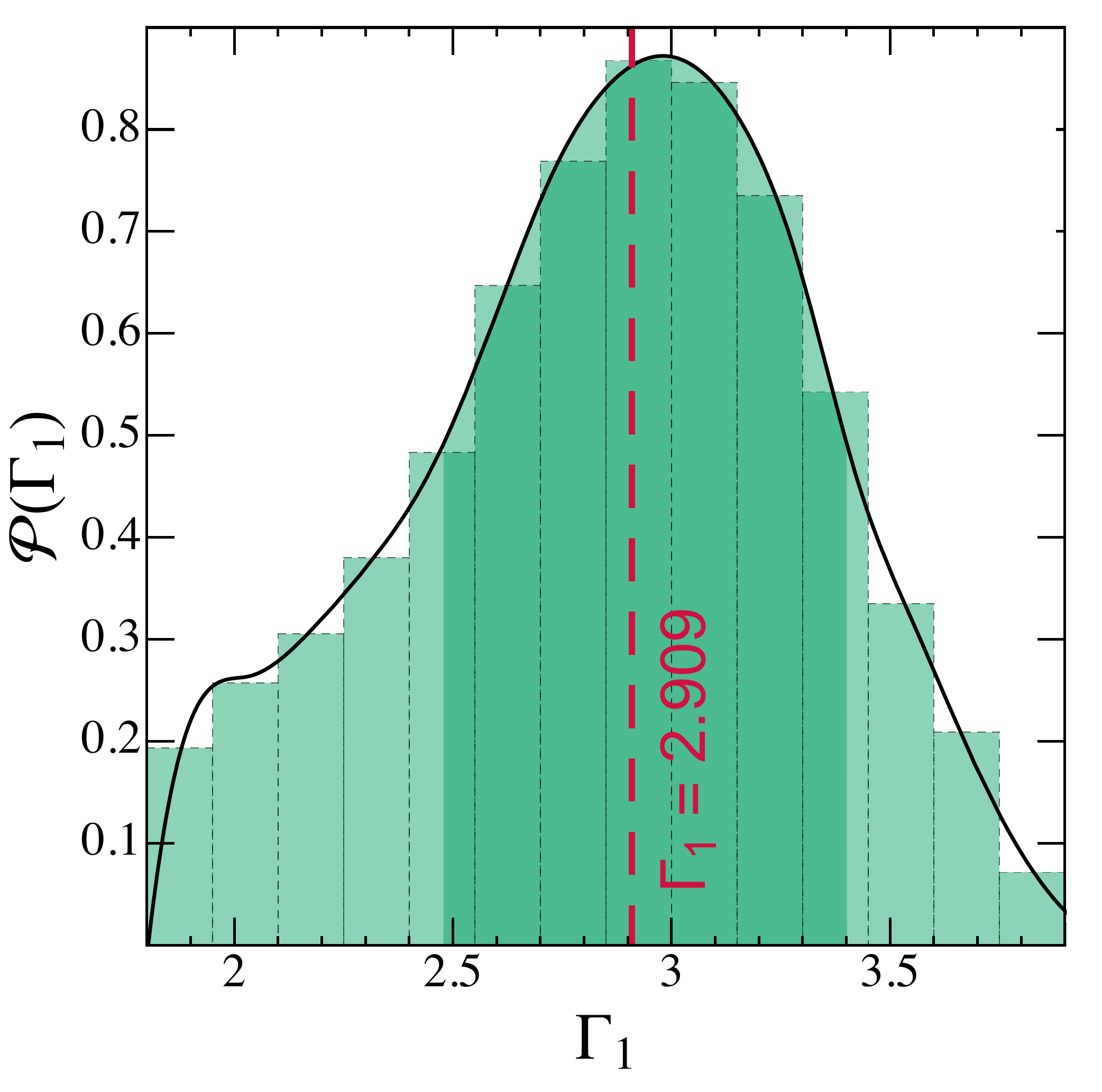}
\includegraphics[width=5.32cm]{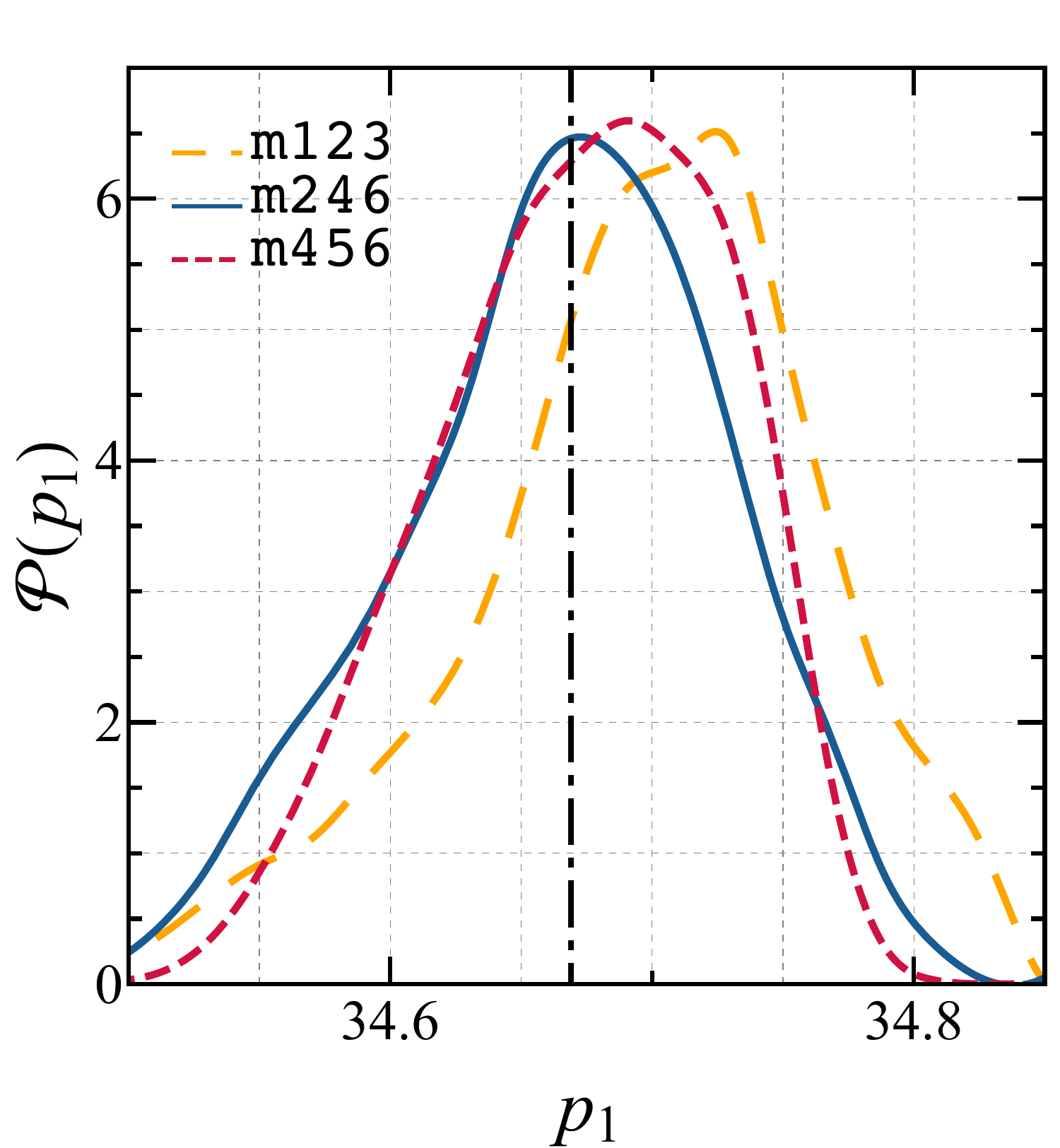}
\caption{(Left and center) Same as Fig.~\ref{fig:Love} but for the
\texttt{h4} EoS.
(Right) We  compare the posterior probability of $p_1$
for  the models \texttt{m246}, \texttt{m456}, \texttt{m123} and the EoS \texttt{h4}.
The vertical dashed line marks the injected value.}
\label{fig:Love2}
\end{figure*}

\section{Reconstruct the EoS parameters}\label{Sec:results1}
The first goal of our approach is to determine the parameters of the
piecewise EoS. As described in Sec.~\ref{Sec:setup} we have 6 unknown
variables to constrain, i.e.,
$\vec{\theta}=\{p_1,\Gamma_1,\Gamma_2,p^c_1,p^c_2,p^c_3\}$, which
require three NS observations. For the sake of clarity, we will test
our method on the following prototype configurations: (i) the model
\texttt{m246} with three objects of mass $(1.2,1.4,1.6)M_\odot$, (ii) a 
\textit{heavier} one \texttt{m456} composed of stars of $(1.4,1.5,1.6)M_\odot$, 
(iii) a \textit{lighter} system \texttt{m123} with masses $(1.1,1.2,1.3)M_\odot$.      

Figure~\ref{fig:Love} shows the marginalized posterior distributions
of the piecewise parameters corresponding to the \texttt{apr4} EoS,
derived for \texttt{m246}. The dashed vertical line in each panel
indicates the true value of the parameter, while the darker bands
correspond to the $1\sigma$ credible intervals. The numerical
values of injected and reconstructed parameters are also listed in
Table~\ref{table:injected} for the considered configurations and 
for the EoS \texttt{apr4} and \texttt{h4}.

At a first glance, we immediately note that the true values of all the
parameters are always reconstructed within the $1\sigma$ confidence
level. The posteriors of the NSs' central pressures are always peaked
around the injected values with nearly symmetrical distributions. The
dividing pressure $p_1$ is also extremely well measured, with the
relative difference between  the expected valued and the median being
below 1\%. 

In general, the adiabatic indices of the piecewise
representation are determined with less accuracy, although some
differences do exist between the various polytropic segments. The top
panels of Fig.~\ref{fig:Love} show indeed that $\Gamma_1$ is
unconstrained, with an almost flat posterior  within the allowed range
of values. 
Conversely, the second index $\Gamma_2$ provides better
results, with a median close to the true quantity, and a probability
distribution that tends to favor larger values. Analyzing the joint
distribution between various parameters we find that $p_1$-$\Gamma_2$
is the only pair that shows a significant correlation, which is,
otherwise, very small.  As we shall see in the next section, this
feature is crucial in order to exploit the piecewise representation to
distinguish different EoS.

Most of the features described so far do not change qualitatively if
we analyze the other two models \texttt{m456} and \texttt{m123}, for
the same EoS \texttt{apr4}.
Smaller masses lead in general to stronger constraints. This is
somehow expected since, for a fixed EoS, lighter NSs yield larger Love
numbers which enhance the tidal contribution to the GW signal and
therefore provide smaller errors $\sigma_\lambda$.  

A direct
comparison between the posterior distribution of $p_1$ and $\Gamma_2$
obtained for the three configurations we consider, is shown in
Fig.~\ref{fig:Love3}. In both panels the best results occur for the
model \texttt{m123}, which is composed of three NSs with masses $(1.1,1.2,1.3)M_{\odot}$, and
the shape of the distribution is similar to that of \texttt{m246}.
Conversely, for \texttt{m456} which 
considers a collection of data with
heavier objects, the
posterior distributions of both $p_1$ and $\Gamma_2$ broaden
significantly and the $1\sigma$ level becomes much looser.

The picture described above changes qualitatively when we consider NSs
made of  a stiffer EoS, which leads to more deformable objects. As an
example, in Fig.~\ref{fig:Love2} we show the probability distributions of the
piecewise parameters for the model \texttt{m246}, assuming
\texttt{h4} as the underlying equation of state. 
We do not plot the central pressures  $p^c_i$
as they are not of great interest in our analysis, although they
are found with an accuracy comparable with that shown in
Fig.~\ref{fig:Love}. 
The left panel of the figure shows that the dividing pressure $p_1$
is, again, the parameter which is constrained with the largest
precision, the posterior distribution being nearly Gaussian and
symmetric around the true value. 
However, a direct comparison with Fig.~\ref{fig:Love} shows that
the role of the adiabatic indices $\Gamma_1$ and $\Gamma_2$ seems now
to be reverted. Indeed, for the EoS \texttt{h4} it is $\Gamma_1$ 
which is   very well estimated, with a relative
difference of the median with respect to the true value smaller than $1\%$.  
The parameter $\Gamma_2$ is essentially unbounded, with a posterior
distribution which is nearly flat. 

\begin{figure}[ht]
\centering
\includegraphics[width=5.7cm]{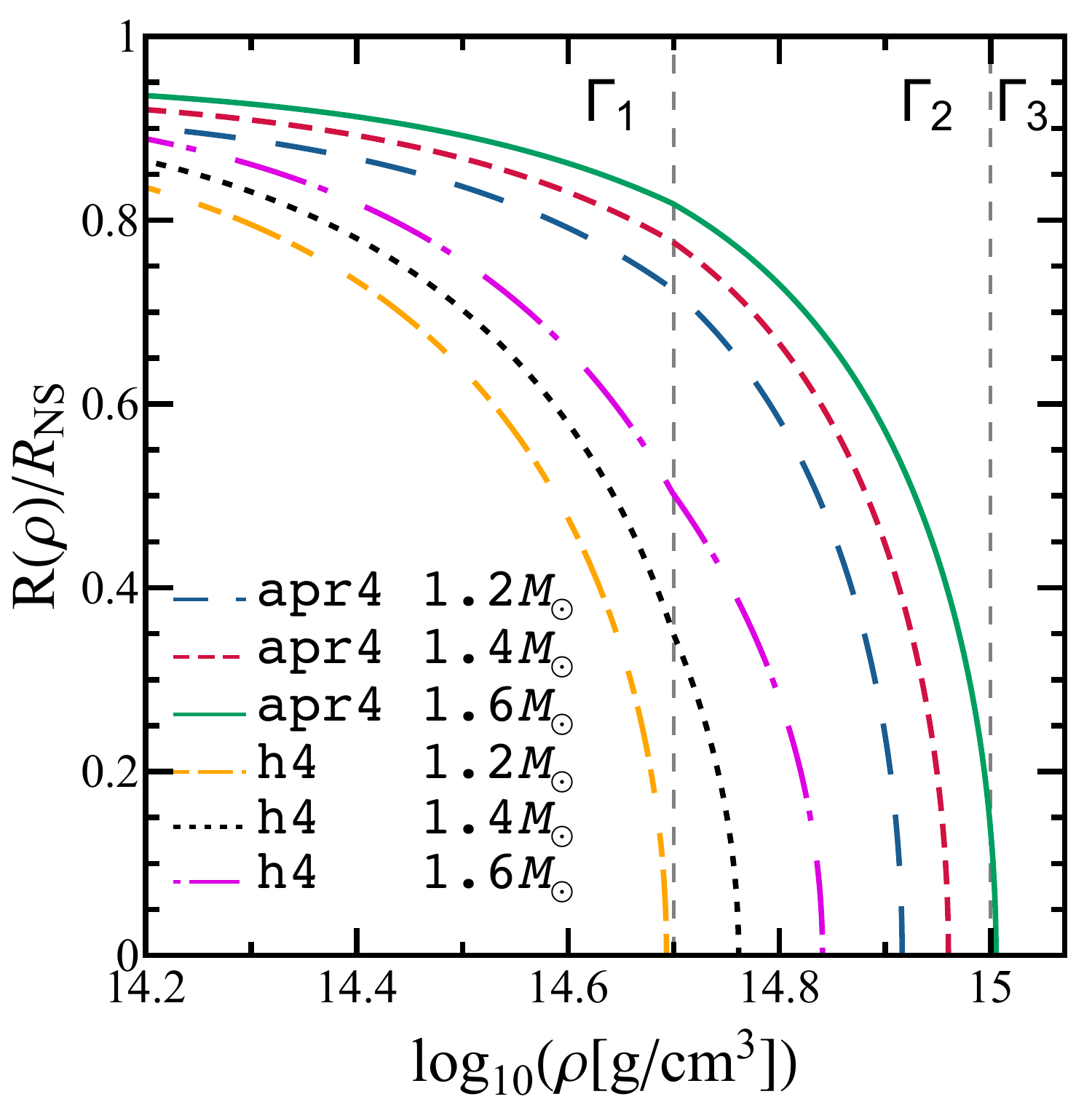}
\caption{The radial distance from the center of the star, normalized
to its radius, is plotted as a function of density.  Different curves
and colors correspond to the masses and EoSs analyzed in the paper.
The vertical lines separate the three regions of the piecewise
parametrization.} \label{fig:radiusprofile} \end{figure}

The different features of the results for the two EoSs can be
understood by looking at Fig.~\ref{fig:radiusprofile}, where we plot,
for each NS and EoS considered for \texttt{m246}, the radial distance $R(\rho)$
normalized to the radius of the star, as a function of the density
$\rho$.  The  major difference between the two EoSs is that the radial
profiles of the \texttt{apr4} stars extend to larger values of $\rho$,
well inside the region of the second branch of the piecewise
polytropic specified by $\Gamma_2$; conversely, the  \texttt{h4} stars
are mainly dominated by the first branch specified by $\Gamma_1$. For
this EoS,  NSs with masses below $1.2M_\odot$ have a central pressure
smaller than $p_1$, and therefore are outside the $\Gamma_2$ interval (see
Fig.~\ref{fig:piecewise}).

Figure \ref{fig:radiusprofile} also shows that at the
boundary between the first two regions, the function $R(\rho)$
of the \texttt{apr4}
stars is already between $80\%$ and $90\%$ of its overall value.
Therefore, it seems quite natural that for this EoS the Love number, which is
proportional to $R_\tn{NS}^5$, is more sensible to variations of $
\Gamma_2$. 
Conversely, the radius of the \texttt{h4} stars  is almost completely
determined by the integration of the stellar equations within the
density region belonging to the first polytropic branch, 
and this is why  the inverse stellar problem constrains
$\Gamma_1$ with a larger accuracy. 

This picture is strongly enhanced for low mass NSs, as can be noticed comparing the right panel of
Fig.~\ref{fig:Love2} and the left panel of Fig.~\ref{fig:Love3},
where we plot ${\cal P}(p_1)$ for the EoS \texttt{h4} and \texttt{apr4},
respectively. For the lightest configuration \texttt{m123}, the MCMC is able to accurately recover the value of $p_1$
for \texttt{apr4}. On the other hand, the reconstructed value for \texttt{h4} shows an offset with respect to the injected parameter.

This result can be traced back  again
to Fig.~\ref{fig:radiusprofile}, which shows that for a $1.2M_\odot$
star, the radius depends only weakly on the adiabatic index
$\Gamma_1$, and it is dominated by the contributions coming from the
low density part of the EoS. In particular, sampling the parameter
space, we have found that the subspace $p_1$-$\Gamma_1$ is
characterized by a large region in which the posterior distribution
assumes values only slightly lower than the absolute maximum, making
extremely difficult to resolve it through the Monte Carlo simulation.
As a consequence, the marginalized distributions are shifted with
respect to the injected values.

\begin{figure*}[th]
\centering
\includegraphics[width=6.2cm]{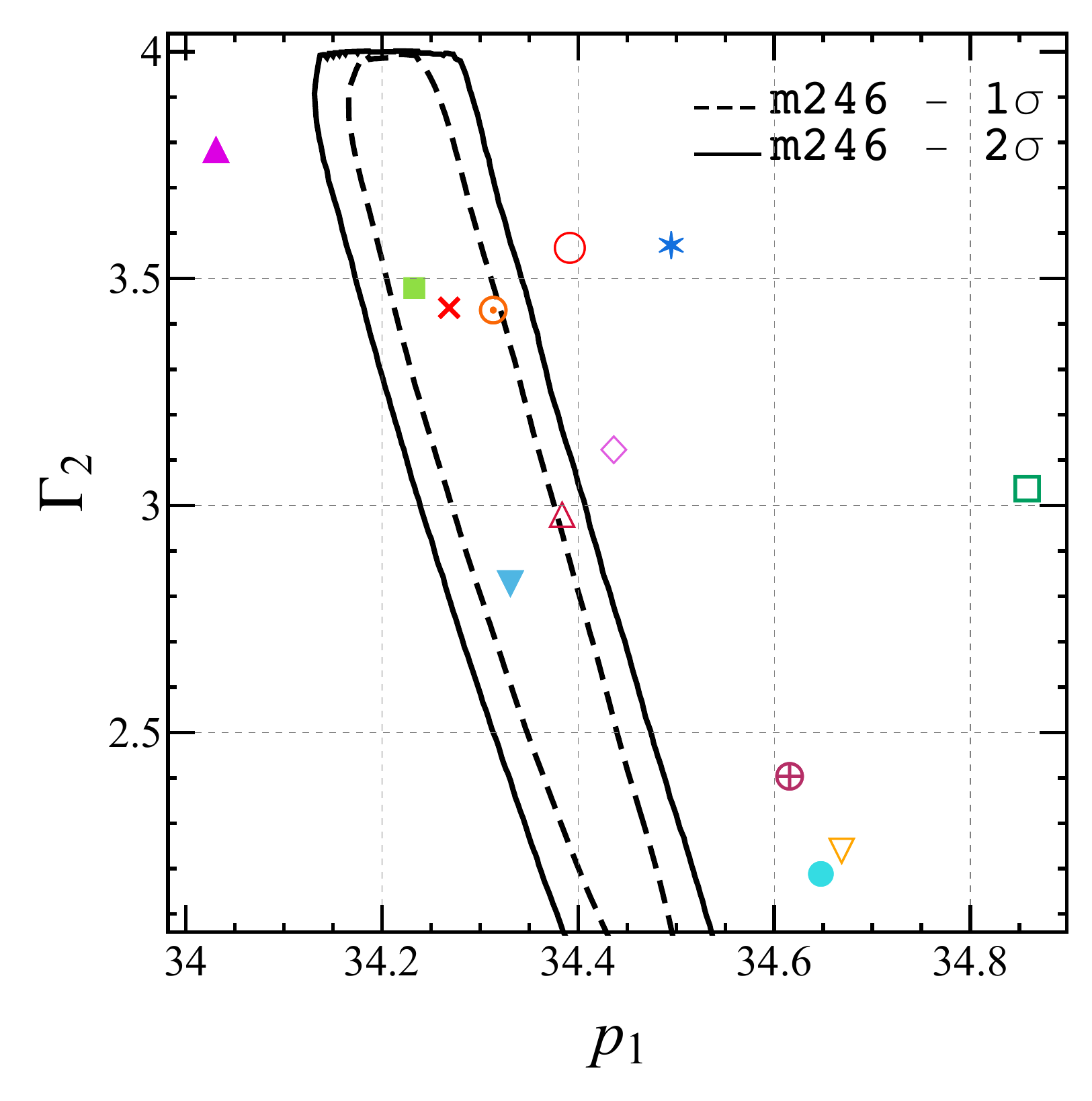}
\includegraphics[width=8.75cm]{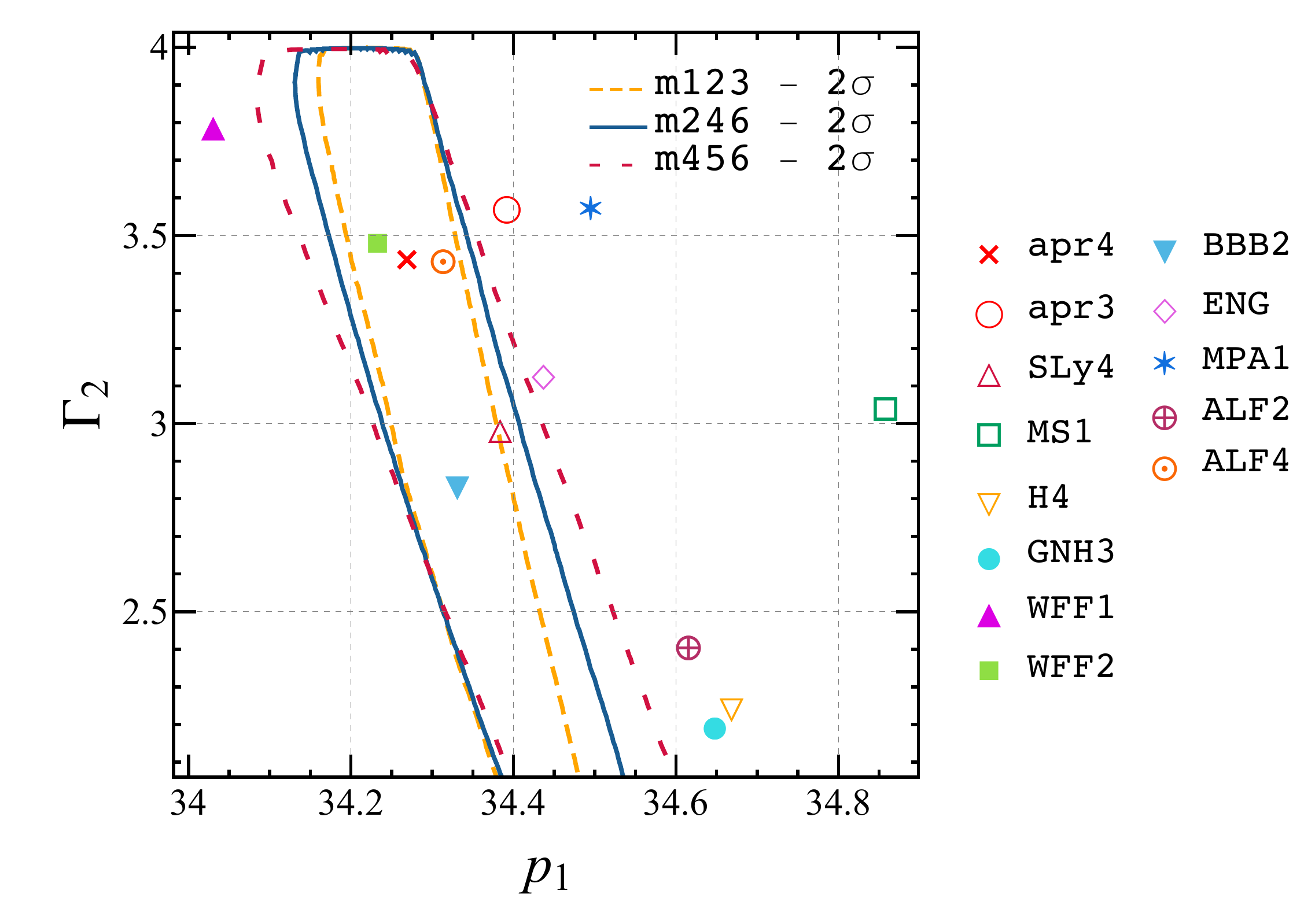}
\caption{(Left) 2-D credible regions at 1- and 2$\sigma$ level for the joint probability distribution $\mathcal{P}(p_1,\Gamma_2)$, computed 
assuming \texttt{apr4} as the true equation of state (red cross), for the model \texttt{m246}. Different markers correspond 
to the values of $p_1$ and $\Gamma_2$ for various EoS. (Right) Comparison among the 2$\sigma$ intervals of $\mathcal{P}(p_1,\Gamma_2)$ derived for the three models considered in this paper.}
\label{fig:mass_comp1}
\end{figure*}

\begin{figure*}[th]
\centering
\includegraphics[width=6.2cm]{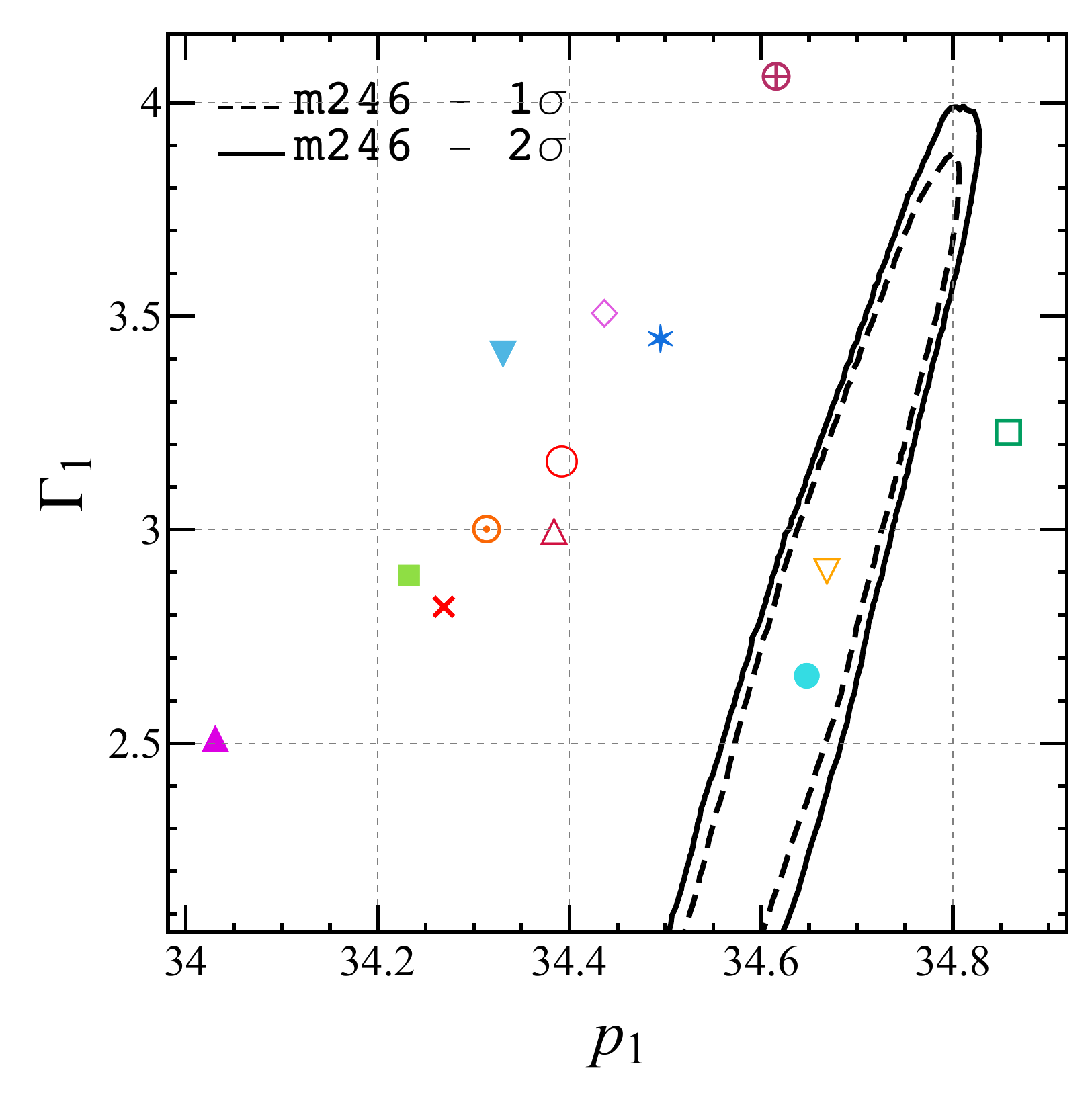}
\includegraphics[width=8.75cm]{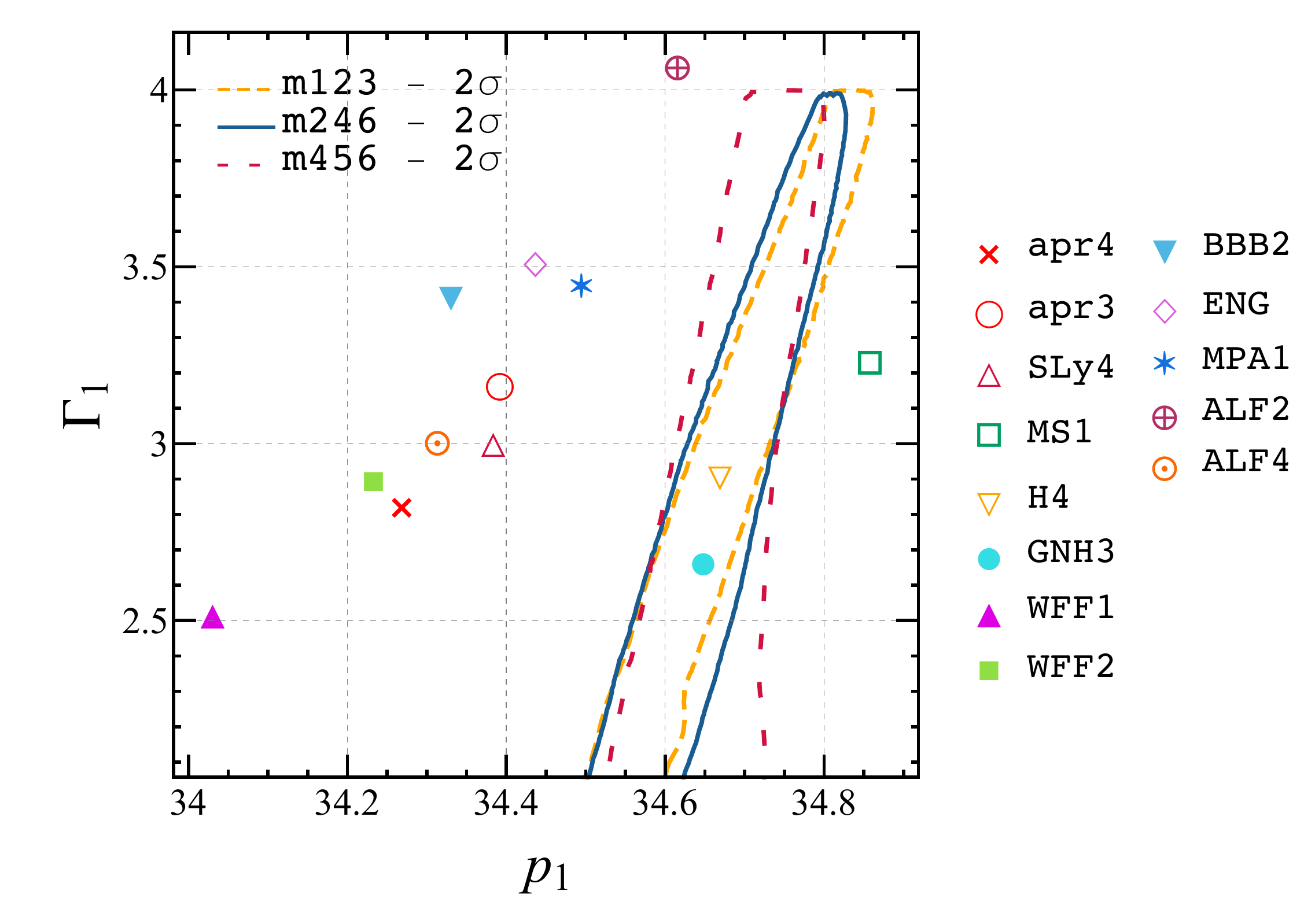}
\caption{(Left) 2-D credible regions at 1- and 2$\sigma$ level for the joint probability distribution $\mathcal{P}(p_1,\Gamma_1)$, for \texttt{h4} as the true equation of state (yellow triangle), for the model \texttt{m246}. Different markers correspond to the values of $p_1$ and $\Gamma_1$ for various EoS. (Right) Comparison among the 2$\sigma$ intervals of $\mathcal{P}(p_1,\Gamma_1)$ derived for the three models considered in this paper.}
\label{fig:mass_comp2}
\end{figure*}

The relativistic inverse stellar problem provides a powerful
framework to perform EoS selection, i.e., to rule out models which are
incompatible with astrophysical observations. 
Most notably, it provides a
straightforward method to combine measurements with different NS
masses and avoiding the quest for approximations which relates
$\lambda$ and $M$ \cite{PhysRevLett.111.071101}.
Our study shows  that for soft (stiff) matter, the joint probability
distribution of $p_1$-$\Gamma_2$ ($p_1$-$\Gamma_1$) offers the best
prospects for EoS selection. To better clarify this statement,
in the left panels of
Figs.~\ref{fig:mass_comp1}-\ref{fig:mass_comp2} we show, for the
configuration \texttt{m246}, the 1- and
$2\sigma$ credible regions obtained from the posterior distributions
of the parameters $p_1$-$\Gamma_2$   for \texttt{apr4},
and $p_1$-$\Gamma_1$ for \texttt{h4}, respectively.
The red cross indicates the injected value, whereas the
different markers are the values of the
parameters corresponding to  various equations of state which have
been mapped on the piecewise polytropic in
\cite{read2009}. 

For both EoSs, the joint distributions seems quite effective in
selecting the correct EoS.  If the true EoS of supranuclear matter is
stiff, measuring the Love numbers with sufficient accuracy would allow
us to essentially rule out almost all known EoSs at more than
$3\sigma$ level. If the
true EoS is  soft, being more similar to \texttt{apr4}, our ability
would worsens, although we may still be able  to constrain a 
portion of the parameter space.  The right panels of
Figs.~\ref{fig:mass_comp1}-\ref{fig:mass_comp2} show how these
bounds slightly change for the various mass configurations which  we have analyzed.

\section{Conclusions}\label{Sec:summary}

The detection of GW170817 opens the possibility to study the behavior of matter at
supranuclear densities using gravitational waves as a probe. Indeed,
even with a single event the LIGO-Virgo
collaboration has set a constraint on the average tidal
deformability $\tilde{\Lambda}$, an EoS-dependent
parameters which appears in  the GW signal, that favours high compact stars.
In the next future, the detection of more events, possibly with
larger signal-to-noise ratio, will sensibly 
strengthen our capability to constrain the NS equation of state
with increasing accuracy.

In this work we have presented a Bayesian approach to reconstruct the
parameters which characterize the EoS in the neutron star core, 
using  masses and tidal Love numbers obtained from GW detections. 
We have developed our analysis modeling the EoS  with a piecewise
polytropic, and generating the mock data 
using a \emph{soft} and a \emph{stiff} EoS which, compatibly with the
constraint already put by GW170817, encompasses the range of admissible
equations of state of nuclear matter.

Our results show that three observations of coalescing neutron star
binaries, by a network of four advanced interferometers--two for LIGO,
plus Virgo and KAGRA--would be sufficient to set interesting
constraints on the parameters of the piecewise polytropic.

The true values of the parameters are always reconstructed  within
$1\sigma$ credible intervals. The
parameter which is determined with the largest accuracy is
the pressure $p_1$, which identifies the interface between the first
and second polytropic branch (see Fig.~\ref{fig:piecewise}). 
In the most favorable scenario, the
error can be smaller than $1\%$.  

On the other hand, bounds on the
polytropic indices are strongly affected by the stiffness  of
the EoS of the mock data. 
If the EoS is soft (stiff) the  smaller error is obtained for
the parameter which characterizes the  inner (outer) part of the core. 
We also find that the central
pressures of the NSs are always determined with an accuracy of the order of
$10\%$. 

Constraints on different parameters can be used to make EoS selection.
In particular, we have found that the joint-2D posterior distribution
for $p_1-\Gamma_2$ (for soft matter) or $p_1-\Gamma_1$ (for stiff
matter) is the best tool to rule out EoS not in agreement with
GW observations.

The method presented in this paper can easily be generalized in
several directions by: (i) including a larger set of masses and tidal
Love numbers obtained from multiple GW events, (ii) combining
different NS observables obtained from astrophysical observations in
the electromagnetic waveband and GW data, (iii) comparing the various phenomenological
EoS available in literature in order to find the model which leads to
the most accurate constraints.  The second point is of particular
interest, since parametrized EoSs are the straightforward
approach to develop multimessenger strategies. A detailed analysis of
how spectroscopic observations of NS radii may be combined with GW
data is already under investigation, and will be presented in a
forthcoming publication \cite{paper2}.

\section*{Acknowledgements}
T.A. was partial supported by the ``NewCompStar,'' COST Action MP1304. T.A. also thanks Kostas Kokkotas for the kind hospitality during his stay at the University of 
T\"{u}bingen. A.M. acknowledges financial support provided under the European Union’s H2020 ERC Consolidator 
Grant “Matter and strong-field gravity: New frontiers in Einstein’s theory” grant agreement no. MaGRaTh–646597.

\appendix

\section*{Appendix: The proposal matrix}\label{Sec:appA}
According to the algorithm GaA, the covariance matrix
$\mathbf{\Sigma}$ of the proposal distribution
$f(\vec{\theta}_1,\vec{\theta}_2)$ is defined as\begin{equation}
\mathbf{\Sigma} = \big(\rho \mathbf{Q}\ \big) \big( \rho
\mathbf{Q}^{\mathrm{T}} \big) \end{equation} where $\rho$ is the step
size of the algorithm and $\mathbf{Q}$ the square root of the
covariance matrix, normalized such that $\mathrm{det}(\mathbf{Q})=1$
\cite{5586491, 1085030}. We compute $\mathbf{Q}$ from
$\mathbf{\Sigma}$ using the Cholesky decomposition.

\renewcommand\algorithmicthen{}
\renewcommand\algorithmicdo{}
\begin{algorithm}[H]
\caption{Adaptive Metropolis-Hastings}\label{mh}
\begin{algorithmic}[0]
\State \textbf{Start}: $\vec{\theta}_1,\rho=1,\mathbf{\Sigma} = \mathbb{1}$
\For{$i=1,\dots,n$}
\State \emph{evaluate} $\mathbf{Q}$ by Cholesky decomposition of $\mathbf{\Sigma}$
\State \emph{normalize} $\mathbf{Q} \to \mathbf{Q}/\mathrm{det}(\mathbf{Q})^{1/D}$ 
\State \emph{propose move} $\vec{y} = \vec{\theta}_i+ \rho \, \mathbf{Q} \cdot \vec{\eta}$ with $\vec{\eta}\sim\mathcal{N}(\vec{0},\mathbb{1})$
\State \emph{evaluate ratio} $\mathcal{P}(\vec{y})/\mathcal{P}(\vec{\theta}_i)$
\If{\textbf{accepted}}
\State $\vec{\theta}_{i+1}= \vec{y}$
\State $\rho \to f_e \, \rho$
\State $\mathbf{\Sigma}  \to \left(1-\frac{1}{N_\tn{C}} \right) \mathbf{\Sigma} 
+ \frac{1}{N_\tn{C}} \big(\vec{\theta}_{i+1} -\vec{\theta}_i \big) \big( \vec{\theta}_{i+1} -\vec{\theta}_i \big)^{\mathrm{T}}$
\EndIf
\If{\textbf{rejected}}
\State $\vec{\theta}_{i+1}= \vec{\theta}_i$
\State $\rho \to f_c \, \rho$
\State $\mathbf{\Sigma}  \to  \mathbf{\Sigma} $
\EndIf
\State \ 
\EndFor
\end{algorithmic}
\end{algorithm}

\begin{figure*}[th]
\centering
\includegraphics[width=18cm]{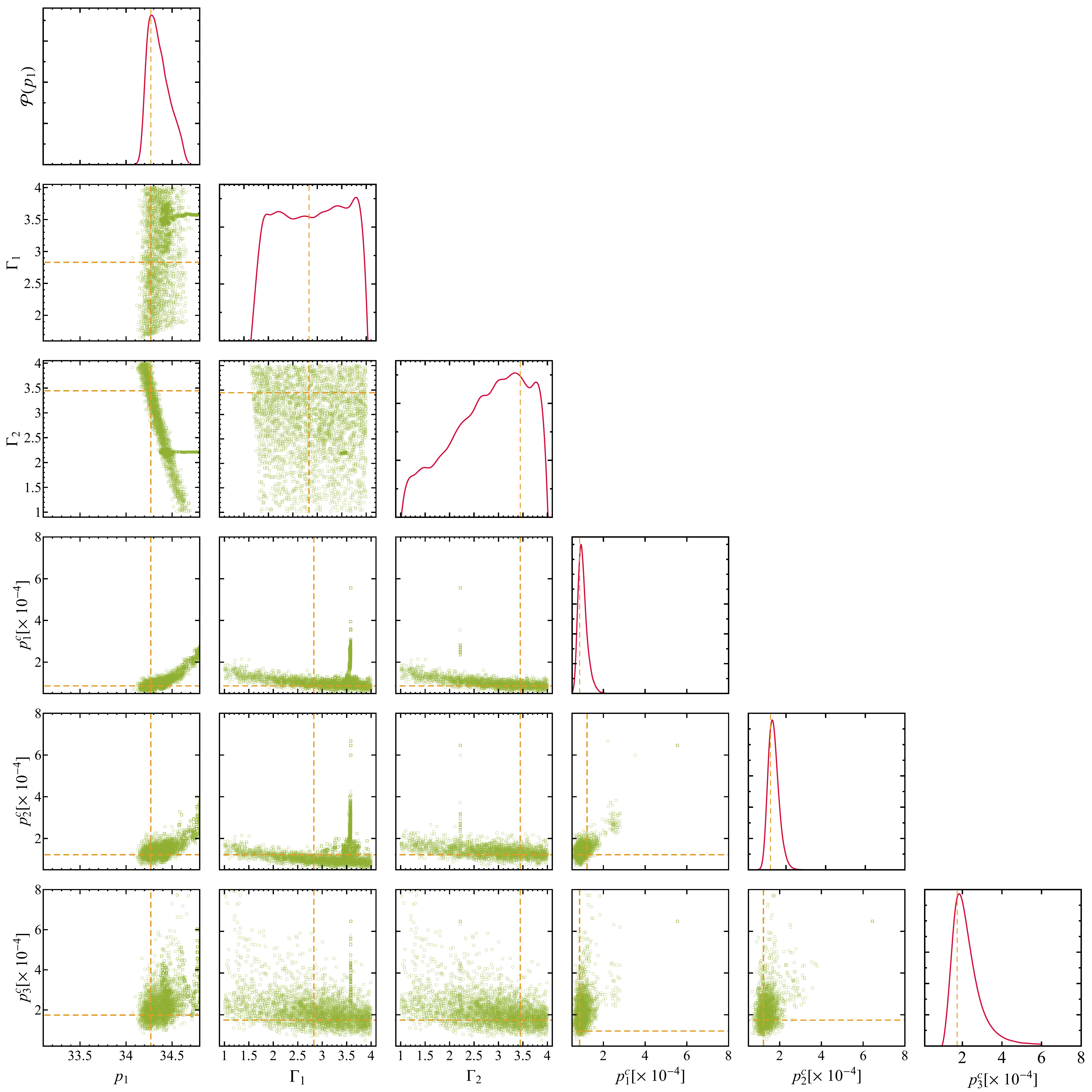}
\caption{An example of the chains produced by the GaA for the model \texttt{m246} and the EoS \texttt{apr4} . On the main diagonal we show the 
marginalized probability distribution of each parameter. }
\label{fig:chain}
\end{figure*}

The structure of the adaptive Metropolis-Hastings algorithm used in
the MCMC is the following: we start from an initial state
$\vec{\theta}_1$, setting $\rho=1$ and $\mathbf{\Sigma} = \mathbf{Q} =
\mathbb{1}$, where $\mathbb{1}$ is the identity matrix. Then, at each
step a new point is sampled as \begin{equation} \vec{\theta}_{i+1} =
\vec{\theta}_{i} + \rho \, \mathbf{Q} \cdot \vec{\eta}\ ,
\end{equation} where $\vec{\eta}$ is drawn from a Gaussian
distribution with zero mean and unit variance
$\mathcal{N}(\vec{0},\mathbb{1})$.  If the proposed move
$\vec{\theta}_{i+1}$ is accepted, the step size and the covariance
matrix are updated according to the following rules: \begin{align}
\rho & \to f_e \, \rho \ ,\\ \mathbf{\Sigma}  \to
\left(1-\frac{1}{N_\tn{C}} \right) \mathbf{\Sigma} &+
\frac{1}{N_\tn{C}} \big(\Delta \vec{\theta} \big) \big( \Delta
\vec{\theta} \big)^{\mathrm{T}} , \end{align} where $f_e>1$ is called
{\it expansion factor}, $N_\tn{C}$ is a free parameter of the GaA and
$\Delta \vec{\theta} =\vec{\theta}_{i+1}-\vec{\theta}_i$. Conversely,
if the proposed jump is rejected, the covariance matrix is not updated
and the step size is reduced by a {\it contraction factor} $f_c<1$:
\begin{equation} \rho  \to f_c \, \rho \quad\ , \quad \mathbf{\Sigma}
\to  \mathbf{\Sigma} \, .  \end{equation} The GaA algorithm relies on
some free parameters, which following \cite{5586491}, we have fixed
to the following values: \begin{equation} \begin{aligned} f_e & = 1+
\beta (1-P) \\ f_c & = 1- \beta P \\ \beta & = 1/N_\tn{C} \\ N_C & =
(D+1)^2 / \log{(D+1)}, \end{aligned} \end{equation} where $D$ is the
dimension of the MCMC parameter space and $P$ is the acceptance
probability of the proposed move. For our simulations we find an
optimal value of such probability, which guarantees an efficient
mixing of the chains, corresponding to $P=0.25$. An example of the 
chain generated with this algorithm for the model \texttt{m246} and the EoS \texttt{apr4} is shown in 
Fig.~\ref{fig:chain}.
\ \\
\ \\
\ \\
\ \\
\ \\
\ \\
\ \\
\ \\

\bibliography{bibnote}

\end{document}